\documentclass[12pt]{article}
\usepackage[utf8]{inputenc}

\title{Quantum undular bores, rainbows, and event horizons in superfluid dam breaks}

\author{L.M. Farrell$^1$, W. Kirkby$^{2,3}$, A. Harris$^1$, D. Tyler$^1$, M. Olshanii$^{4}$, \\ D.H.J. O'Dell$^{1,\ast}$}

\date{$^1$Department of Physics and Astronomy, McMaster University, 1280 Main St. W., Hamilton, ON, Canada L8S 4M1\\
$^2$ Kirchhoff-Institut für Physik, Universität Heidelberg, Heidelberg, Germany\\
$^3$ Physikalisches Institut, Universität Heidelberg, Heidelberg, Germany\\
$^4$ Department of Physics, University of Massachusetts Boston, Boston, MA 02125, United States of America\\
      [1ex]
\textsuperscript{*}Corresponding author. E-mail: dodell@mcmaster.ca\\[2ex]
    \today}

\usepackage[
  backend=biber,
  style=numeric-comp,
  sorting=none,
  sortcites=true
]{biblatex}

\usepackage{float}
\usepackage{hyperref}
\usepackage{graphicx}
\usepackage{amsmath}
\usepackage{amssymb}
\usepackage{hyperref}
\usepackage{caption}
\usepackage{xcolor}
\usepackage{dsfont}
\usepackage[margin=2.5cm]{geometry}

\begin{document}

\maketitle
\begin{abstract}
The sudden removal of a potential barrier from a Bose-Einstein condensate (BEC) gives rise to a quantum version of a hydrodynamic dam break and leads to rich wave dynamics that have similarities to other localized defect problems such as domain wall dynamics in spin systems.  Denoting $\Delta n$ as the initial difference in density between the upper $n_{1}$ and lower $n_{0}$ reservoirs on either side of the dam, we use the Gross-Pitaevskii equation to study the quasi-one dimensional case in both perturbative ($\Delta n \ll n_{0}$) and non-perturbative ($\Delta n \sim n_{1}$) regimes. In the perturbative regime a pair of outwardly propagating dispersive wavepackets forms which can be viewed as quantum versions of undular tidal bores that have analytic forms at long times in terms of the integrals of Airy functions with a wavelength that grows as $(\hbar^2 t)^{1/3}$. Airy functions are the universal wave functions that dress structurally stable fold caustics where pairs of rays coalesce, and, indeed, we show that the quantum dam break problem has the same ray structure as the naturally occurring caustic phenomenon of a double rainbow, including Alexander's dark band between the two bows where no light is scattered: we identify the intermediate density plateau in the dam break as an analogous `silent band' where only evanescent sound waves can exist. In the opposite regime of a non-perturbative dam break we show that a self-induced sonic horizon occurs when $\Delta n \geq (8/9)n_{1}$. A discussion of possible experimental schemes for amplification of quantum undulations is included as well as an alternative scheme for dispersive wave generation where a constant flow is imprinted on a BEC in box trap.
\end{abstract}

\section{Introduction}

Experiments on expanding Bose-Einstein condensates (BECs) can be used to study a variety of quantum many-particle dynamics, ranging from exploding `bosenovas' \cite{Donley_2001}, Heisenberg-limited momentum spread \cite{Gotlibovych_2014}, and emergent generalized hydrodynamics out-of-equilibrium \cite{Schemmer_2019,Bouchoule_2023,Dubois_2024}, to simulating the behaviour of quantum fields in an expanding universe \cite{Fedichev_2004,Uhlmann_2005,Jaskula_2012,Eckel_2018,Llorente_2019,Banik_2022,Bhardwaj_2024}. In this work we analyze the expansion of quasi-one dimensional (1D) BECs, which have been realized experimentally using quantum gases in waveguides, optical lattices, and highly anisotropic traps \cite{Gorlitz_2001,Bongs_2001,Strecker_2002,Morsch_2002,Esteve_2006,Muller_2008,Fang_2010,Armijo_2011,Hamner_2011,Ryu_2015,Navez_2016,Everitt_2017,Nguyen_2017}, as well as in polariton BECs \cite{Falque_2025}. In particular, motivated by classical hydrodynamic dam breaking problems \cite{Hunt_1984,Stoker_1992,Chanson_2004,Feria_2006,Ostapenko_2007,Korobkin_2009,Dutykh_2010,Castro_2017}, we study dispersive waves (DWs) that are generated in a superfluid version of a dam break in a quasi-1D BEC.

\begin{figure}[!h]
	\centering
	\begin{minipage}{0.333\textwidth}
		\centering
		\includegraphics[width=5.2cm]{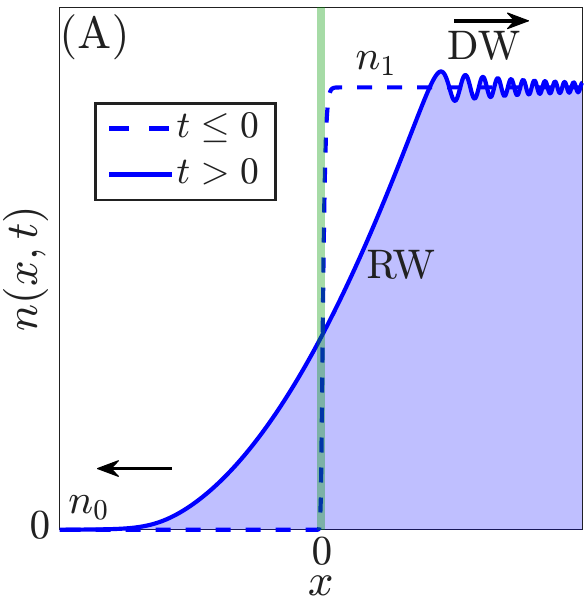}
	\end{minipage}%
    \begin{minipage}{0.333\textwidth}
		\centering
		\includegraphics[width=5.2cm]{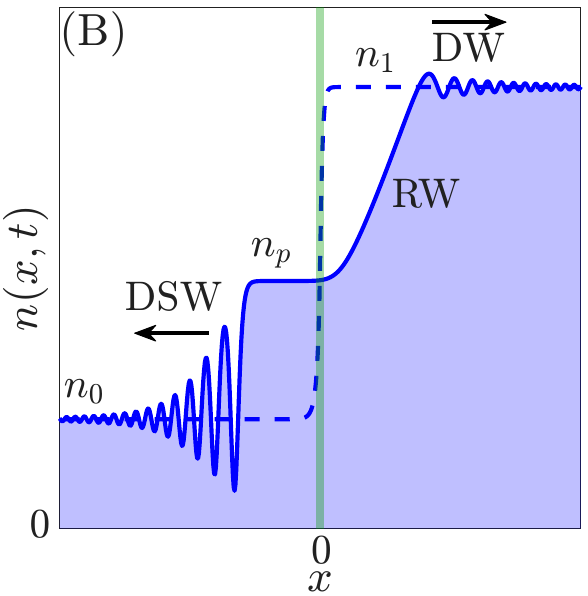}
	\end{minipage}%
	\begin{minipage}{0.333\textwidth}
		\centering
		\includegraphics[width=5.2cm]{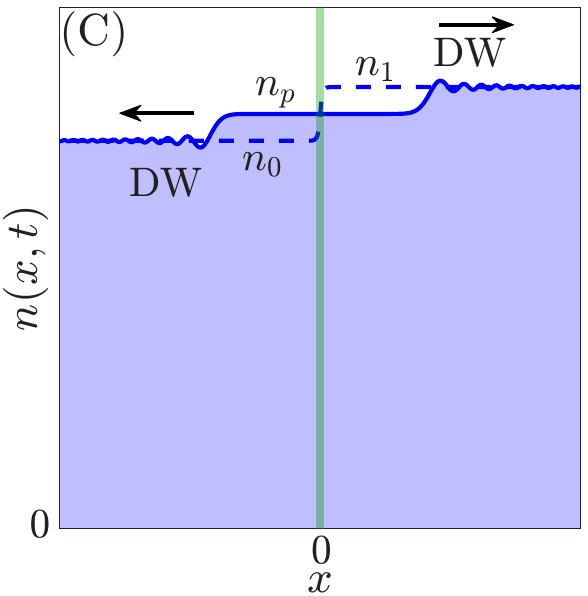}
	\end{minipage}\\[-0.2cm]
	\caption{Three classes of dam breaking within a zero-temperature BEC, obtained numerically via equation (\ref{eq:gpe}): \textbf{(A)} Dry-channel $\Delta n= n_{1}$, \textbf{(B)} Wet-channel $\Delta n \sim n_{1}$, and \textbf{(C)} Perturbative wet-channel $\Delta n\ll n_{0}$. In all cases the condensate starts at rest with an initial step-like density profile (dashed blue lines) due to a steep potential barrier (solid green line) at $x=0$ applied with different strength in the three cases. The resulting density step is softened at the scale of the healing length $\xi$ by quantum pressure. After switching the barrier off at $t=0$ (i.e.\ dam breaking) a snapshot of the resulting condensate dynamics is depicted (solid blue lines). Depending on the situation, DWs, DSWs, rarefaction waves (RWs), and/or an intermediate plateau $n_{p}$ can form. The DW oscillations along the top of the step in \textbf{(A)} and \textbf{(B)}, and at the top and bottom in \textbf{(C)}, are quantum versions of the undulations seen in tidal bores.} \label{fig:damBreakProblems}
\end{figure}

As illustrated in figure \ref{fig:damBreakProblems}, we imagine a setup where a steep potential barrier initially separates a region of higher particle density $n_{1}$ (upper reservoir) from one of lower density $n_{0}$ (lower reservoir). The barrier is suddenly removed and the BEC freely expands in 1D. A key parameter that determines the ensuing wave dynamics is the initial density difference between the reservoirs $\Delta n= n_{1}-n_{0}$, three examples of which are displayed: a dry-channel, a wet-channel, and a perturbative wet-channel.

Although in many ways BECs behave like inviscid classical fluids, they additionally experience `quantum pressure' which comes from the zero-point motion of the atoms and contributes to the total energy in proportion to $\hbar^2$. Quantum pressure is the origin of dispersion in the Bogoliubov excitation spectrum for small amplitude excitations about the condensate, and without it the excitations obey a purely linear dispersion relation $\omega=ck$, where $c$ is the speed of sound in the condensate. When the interaction and trapping energies dominate the quantum pressure energy (as they can, for example, in the ground state of a harmonically trapped BEC), it is common to make the Thomas-Fermi (TF) approximation and neglect the latter \cite{Castin_1996,Pitaevskii_2003,Brazhnyi_2003,Kamchatnov_2004b,Campbell_2015,Ivanov_2019,Jalm_2019,Olshanii_2021,Olshanii_2022,Molinero_2022}. This corresponds to taking the classical limit $\hbar\rightarrow 0$. However, the role of dispersion becomes significant precisely where the TF approximation predicts singularities.

A well-known example of dispersion regulating singularities in fluid dynamics is provided by dispersive shockwaves (DSWs) where non-linearity is balanced by dispersion \cite{Zabusky_1965,Gurevich_1973,El_2016,Kamchatnov_2021}. In the second dam break case shown in figure \ref{fig:damBreakProblems}, a DSW forms in the lower reservoir where the TF solution becomes multi-valued and hence unphysical but is regulated by dispersion leading to a train of solitons. This requires both a sizable value of $\Delta n$ to generate enough non-linearity and of course a non-zero density in the lower reservoir. Superfluids make a good platform for studying DSWs since the dissipation that often masks dispersion can in principle be absent and DSWs have been intensively studied in Bose gases both theoretically \cite{Kamchatnov_2004,Damski_2004,Hoefer_2008,Salasnich_2007,Salasnich_2016,Ivanov_2017,Simmons_2020,Dubessy_2021,Wang_2023,Simmons_2023,Chandramouli_2024,Mohapatra_2026,Yang_2026} and experimentally \cite{Dutton_2001,Simula_2005,Hoefer_2006,Meppelink_2009,Mossman_2018,Mossman_2020,Mossman_2025,Schuttelkopf_2026}, while experiments with superfluid helium films are also advancing \cite{Reeves_2025}. DSWs are also an active area of investigation in a variety of other physical systems including ultracold fermionic gases \cite{Eisler_2013}, plasmas \cite{Ivanov_2020}, some classical fluids \cite{Caputo_2003,Leach_2008,Leach_2010,Needham_2021}, and non-linear optical media \cite{Kodama_1999,Wan_2007,Isoard_2021,Hang_2023}.  
In these studies Whitham's modulation theory is often the theoretical method of choice as it can deal with strong non-linearity \cite{Whitham_1965,Whitham_1974,Fornberg_1978,An_2018,Li_2024,Moller_2024}.  

In this paper, although we shall consider non-perturbative dam breaks numerically, we will focus most of our analytic effort on \textit{singularities that occur even in the perturbative regime} $\Delta n\ll n_{0}$, as shown in figure \ref{fig:damBreakProblems} (C). This corresponds to a region of parameter space where the wave dynamics are linear, see also the superfluid region in figure 1 A of reference \cite{Reeves_2025}. In this regime DWs in BECs are dominated by quantum effects, owing their formation to quantum pressure, whereas the non-perturbative dam break DSWs correspond to the cnoidal/solitary wave type \cite{Reeves_2025} and are no longer purely quantum. Related scenarios have been studied in references \cite{Damski_2004,Salasnich_2016,Wang_2023}.
Our approach will be to adapt a theoretical description used for undular tidal bores in rivers \cite{Berry_2018,Berry_2019} (see also \cite{Kamchatnov_2021}) to the quantum dam break problem. Due to this connection we refer to the DW oscillations seen in figure \ref{fig:damBreakProblems} along the top of the step in panels (A) and (B), and at the top and bottom in (C), as \textit{quantum undulations}. A similar mathematical problem also occurs in domain wall expansion dynamics in spin-chain models \cite{Fagotti_2017,Bulchandani_2019,Wei_2022,Riddell_2023,McRoberts_2024,Fujimoto_2024}, emphasizing the universal nature of the basic phenomenon.

Two features of dam breaks that we highlight in this paper are caustics and event horizons. Caustics are a wave focusing phenomenon that produces sharp intensity singularities in the ray limit $\hbar \rightarrow 0$. The number of rays changes as we cross a caustic corresponding to bifurcations of the classical solutions. Analyzing the rays underlying the DWs in a perturbative dam break reveals that each wavefront corresponds to a \textit{fold caustic} where two rays coalesce (or two are born). An everyday example of a fold caustic is the optical rainbow \cite{Khare_1974,Berry_2015}. In fact, the back-to-back pair of wavefronts formed in the perturbative dam break pictured in figure \ref{fig:damBreakProblems} (C) have the same ray structure as a \textit{double rainbow}. The analogy also extends to the dark region between the two rainbows where almost no light scatters. This is known as Alexander's dark band after Alexander of Aphrodisias who discussed it around A.D. 200 \cite{Nussenzveig_1977}. In the BEC case we find an equivalent region between the two wavefronts where no sound waves propagate which we therefore refer to as a `silent band'. Furthermore, the wave functions that dress the caustics and resolve the singularities are Airy-type functions in both cases. The same ray and wave structures also occur in dynamical phase transitions following a sudden quench in certain spin models \cite{Link_2024}. This universality is explained by catastrophe theory, which says that fold caustics are \textit{structurally stable} against perturbations and hence occur generically in nature because they do not require fine tuning \cite{Berry_1981,Gilmore_1981,Nye_1999}. In fact, the only structurally stable catastrophe in 1D is the fold. One of the consequences of this universality is that it gives rise to scaling relations such as a $t^{1/3}$ time dependence we find for the wavelength of the DWs.

Caustics also provide a natural framework for discussing the ray bifurcations that occur near black holes and give a novel interpretation of Hawking radiation \cite{Leonhardt_2002,Farrell_2023}. Sonic analogues of event horizons can occur in fluids where the flow speed exceeds the local speed of sound \cite{Unruh_1981,Barcelo_2026}, and tidal bores in rivers turn out to correspond to white holes where waves cannot enter the supercritical region, i.e.\ a time-reversed black hole \cite{Volovik_2005,Volovik_2006}. Although we use a similar mathematical construction as tidal bores, here we show for both dry- and certain wet-channel dam breaks that a sonic analogue of a black hole horizon forms, and specifically one which is dynamic as opposed to stationary \cite{Kolobov_2021,Fabbri_2021,Fourdrinoy_2022,Balbinot_2022}; the dry-channel case was recently investigated experimentally in reference \cite{Sharan_2025}, while the wet-channel case was studied theoretically in reference \cite{Cao_2025}. Our work highlights that the sonic horizon which forms during dry-channel breaking is dynamically simpler than that of wet-channel breaking, and for a certain critical value of $\Delta n_{\mathrm{crit}}$ a horizon of finite extent (width) forms \cite{Penalver_2025,Penalver_2026}. Although Hawking radiation is fundamentally a stationary process \cite{Hawking_1975}, realistic black holes are expected to be dynamic not just during their formation, but during their metastable phase and possible evaporation \cite{Hayward_1994,Ashtekar_2004,Nielsen_2006,Nielsen_2008,Vertogradov_2025}. 

The outline for the rest of this paper is as follows. In Section \ref{sec:quantumhydrodynamics} we briefly introduce the equations of quantum hydrodynamics that stem from the Gross-Pitaevskii equation (GPE). In Section \ref{sec:perturbative} we construct analytic time-dependent solutions to these equations for the case of perturbative dam breaks, treating the resulting wavepackets as quantum versions of the undulations seen in undular tidal bores. In Section \ref{sec:asymptotic} we obtain the $t \rightarrow \infty$ asymptotics of these solutions where they reduce to a universal wave function in the form of the integral of an Airy function that matches the exact GPE numerics even at moderate times. Simple expressions for the amplitude and wavelength are derived. In Section \ref{sec:nonperturbative} we discuss the non-perturbative dam break regime where the symmetry between the lower and uppers DWs is broken; the lower DW becomes larger in amplitude and transitions into a DSW, reflecting the increasing non-linearity in expansion dynamics for larger $\Delta n$. The possibility of experimental observation in the quantum regime is also discussed; by quenching the scattering length at the moment of breaking, the wave amplitude can be increased for an improved detection signal during the subsequent dynamics. In Section \ref{sec:rainbows} we show that the pair of DWs created by the dam break are close analogues to optical double rainbows \cite{Khare_1974,Berry_2015}, and also show that in the TF approximation a sonic event horizons forms when $\Delta n > (8/9) n_{1}$. In Section \ref{sec:imprintedflow} an alternative scheme for generating dam break waves via an initially imprinted flow without the presence of an actual dam is proposed and studied.

\section{Quantum hydrodynamics}
\label{sec:quantumhydrodynamics}

The dynamics of a zero-temperature BEC, consisting of $N$ atoms of mass $m$ trapped in an external potential $V(\mathrm{r},t)$, is governed by the GPE \cite{Pitaevskii_2003,Pethick_2008}
\begin{equation}
    i\hbar\partial_{t}\Psi(\mathbf{r},t) =\bigg(-\frac{\hbar^{2}}{2m}\nabla^{2}+V(\mathbf{r},t)+g\vert\Psi(\mathbf{r},t)\vert^{2}\bigg)\Psi(\mathbf{r},t) \ , \label{eq:gpe}
\end{equation}
where $\Psi(\mathbf{r},t)$ is the condensate wave function normalized by the total number of atoms $N=\int\vert\Psi\vert^2d^{3}r$ such that the local atom density is given by  $n(\mathbf{r},t)=\vert \Psi(\mathbf{r},t)\vert^2$, and $V(\mathbf{r},t)$ describes an external potential. In this paper we take $g>0$, accounting for repulsive short-range (contact) interactions between the atoms, and in 3D is given by $g_{3D}=4\pi\hbar^{2}a/m$ where $a$ the s-wave scattering length. To study dam breaking in 1D, we assume a harmonic trapping potential along the radial direction $V_{\perp}(r)=m\omega_{\perp}^{2}r^{2}/2$, where $r^{2}=y^{2}+z^{2}$, that is tight enough to restrict the dynamics to being purely along the $x$-axis, giving rise to an effectively 1D (cigar shaped) condensate with interaction constant $g=g_{\mathrm{3D}}/(2\pi l_{\perp}^2)$. The characteristic length scale of the harmonic oscillator is $l_{\perp}=\sqrt{\hbar/m\omega_{\perp}}$, and the quasi-1D condition is satisfied when $g n\ll \hbar\omega_{\perp}$. Throughout this work we compare numerical simulations of the GPE obtained via the split-step-Fourier method against our analytic theory using the following realistic choices of parameters \cite{Roati_2007,Duchene_2026}: $N=4\times10^{4}$ potassium-39 atoms, $\omega_{\perp}=2\pi\times 800$ Hz, $a=100 a_{0}$ where $a_{0}$ is the Bohr radius, and the total system length is $L=800$ $\mu$m -- sufficiently large so that DWs are not able to propagate and reach the simulation edges during the timescales we consider. For perturbative dam breaks we set $n_{0}=50$ atoms/$\mu$m and choose $n_{1}$ such that $\Delta n\approx 0.04 n_{0}\ll n_{0}$.

The quasi-1D condensate wave function $\psi(x,t)$ can be usefully expressed in terms of the density $n(x,t)$ and phase $\theta(x,t)$ by $\psi(x,t)=\sqrt{n(x,t)}\textrm{e}^{i\theta(x,t)}$. The superfluid flow velocity is obtained by taking the gradient of the phase, $v(x,t)=\frac{\hbar}{m}\partial_{x}\theta(x,t)$. The GPE can then be shown to be exactly equivalent to a continuity equation and a quantum Euler equation
\begin{gather}
    \partial_{t}n+\partial_{x} (n v)  =  0 \ , \label{eq:cont}\\
    \hbar \, \partial_{t} \theta+ \left(\frac{1}{2}m v^{2}+V+gn-\frac{\hbar^2}{2m \sqrt{n}} \partial_{x}^{2} \sqrt{n} \right)  =  0 \ . \label{eq:euler}
\end{gather}
The only term that differs from those for an idealized 1D inviscid classical fluid is the quantum pressure term proportional to $\hbar^2$ in equation (\ref{eq:euler}) \cite{Stringari_1996,Pitaevskii_2003,Pethick_2008}. The origin of this term is quantum zero point motion of the particles, which is neglected in the TF limit. This limit is applicable when the length scale of density modulations is much larger than the healing length, defined as 
\begin{equation}
    \xi=\hbar/\sqrt{mgn}=\hbar/mc \ ,\label{eq:healinglength}
\end{equation}
 where $c=\sqrt{gn/m}$ is the speed of sound in the BEC. The healing length sets the scale over which seemingly sharp density steps are smoothed, as can be seen in the $t = 0$ density profiles in figure \ref{fig:damBreakProblems}. The healing length is analogous to the Compton wavelength of relativistic quantum mechanics with the speed of sound replacing the speed of light.

Quantum pressure plays an important role in the phenomena we investigate. Apart from limiting the maximum steepness of the initial density profile, it is responsible for the dispersive behaviour of waves excited in the dam break. This is evident from the Bogoliubov dispersion relation for elementary excitations (quasi-particles) in a uniform BEC \cite{Pitaevskii_2003}
\begin{equation}
    \omega_{\pm}(k)=\pm \sqrt{c^2 k^2+\frac{ \xi^2 c^2 k^4}{4}}  \ . \label{eq:bogDispFluid}
    \end{equation}
This is non-linear such that waves with $k \gtrsim 1/\xi $ are supersonic, and the non-linearity is controlled by $\xi^2 \propto \hbar^2$. In accordance with the analogy to relativity, $\omega_{\pm}$ in equation (\ref{eq:bogDispFluid}) can be associated with particles and antiparticles, respectively (in condensed matter applications $\omega_{\pm}$ are associated with particle and hole excitations). Both $\omega_{+}$ and $\omega_{-}$ waves are included in the description of the excitations created by the dam break given in Section \ref{subsec:dispersive_analytical_model} below. 

Since we shall later apply the mathematical solutions for the waves created in undular tidal bores to the case of dam breaks, it is interesting to compare the Bogoliubov dispersion relation with that for long waves on the surface of shallow water which is used to describe such bores  \cite{Berry_2018}
\begin{equation}
\Omega_{\pm}(k)= \pm \sqrt{g_{\mathrm{grav}} \, k \, \tanh[d_{r} \, k]} \ .
\label{eq:waterdispersion}
\end{equation}
In this expression $g_{\mathrm{grav}}$ is the acceleration due to gravity and $d_{r}$ is the depth of the river. The speed of these waves at long wavelengths is $\sqrt{g_{\mathrm{grav}} \ d_{r}}$ and the length scale that plays a role analogous to that of the healing length in BECs and hence controls the dispersion is now $d_{r}$. The shallow-water  dispersion relation bends in the opposite direction to (\ref{eq:bogDispFluid}) so that waves with $k > 1/d_{r}$ are slower than those at $k=0$.

\begin{figure}[!t]
    \centering
    \captionsetup{width=1\linewidth}
    \includegraphics[width=9cm]{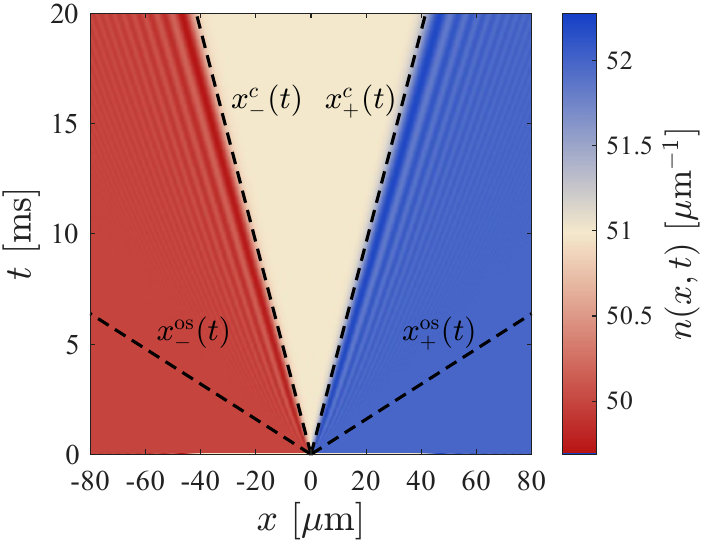}\\[-0.2cm]
    \caption{Sound cone formed in spacetime by the outward propagation of DWs from the site of the dam break at $x=t=0$. We plot the density profile which is obtained by numerically solving the GPE (\ref{eq:gpe}) for the case of a perturbative wet-channel dam break with $n_{0}=50\mu$m$^{-1}$ and $\Delta n\approx 0.04 n_{0}\ll n_{0}$. The outer pair of dashed lines obey $x_{\pm}^{os}(t)=\pm 3(1+\sqrt{n_{p}/n_{0}})c_{0}t$ and mark out the extreme edges of the undulations (where the amplitude of the DW vanishes). These edges propagate faster than the background speed of sound $c_{0}$. The inner pair of dashed lines obey $x_{\pm}^{c}(t)=\pm c_{0}t$ and indicate the outward propagation at speed $c_{0}$ of the DW fronts (defined in the text) immediately preceding the largest amplitude undulations. No waves appear inside the cone [which corresponds to the plateau in figure \ref{fig:damBreakProblems} (C)] and in this sense it is inside-out in comparison to cones in spin chains when a single spin is flipped, where the undulatory part of the wave is inside the cone rather than outside. Similar diagrams can also be found in references \cite{Hang_2023,McRoberts_2024}. }\label{fig:spaceTime}
\end{figure}

Focusing on a perturbative dam break in a BEC like that shown in figure \ref{fig:damBreakProblems} (C), we see that: i) the lower and upper edges of the initial density profile respectively begin moving to the left and right away from the origin, forming a pair of back-to-back approximately symmetric travelling DWs (undulations), and ii) an intermediate plateau of constant density $n_{p}$ forms in between and provides the new equilibrium state of the system. The resulting sound cone that forms due to the spreading of the DWs in space and time is visible in figure \ref{fig:spaceTime}. As expected, the DWs spread faster than the speed of sound determined by the linear part of the Bogoliubov dispersion relation. It is notable that no waves propagate inside the cone and we shall explain in Section \ref{sec:rainbows} how this is a sonic version of Alexander's dark band in optical double rainbows. In fact, this sound cone is inside-out in comparison to cones in related localized defect problems such as domain wall relaxation or when a spin is flipped in a spin chain \cite{Kirkby_2019,Kirkby_2022}, where the undulatory part of the wave is inside the cone rather than outside.

\section{Perturbative wet-channel dam break}
\label{sec:perturbative}

\subsection{Dispersive analytical model}
\label{subsec:dispersive_analytical_model}

We now show that perturbative dam breaks in BECs are closely related to undular tidal bores. The connections between bores and dam breaks have previously been studied by a number of authors, including \cite{Fornberg_1978,Stoker_1992,Chanson_2004,Furuyama_2008,Whitham_1974,El_2016,Castro_2017,An_2018,Castro_2020,Castro_2021,Kamchatnov_2012,Kamchatnov_2021}. Our treatment adapts Berry's simple minimal model for undular tidal bores in rivers \cite{Berry_2018,Berry_2019} 
to the case of a BEC. Its chief feature is a solution based on a linear superposition of DWs that exactly satisfies the boundary conditions at $x=\pm \infty$. In the tidal bore case the waves are assumed to initially emanate from a small step-like perturbation on the back of a deeper river with downstream uni-directional flow. The river depth is taken to be large compared to the initial bore height, which is equivalent to our perturbative condition $\Delta n\ll n_{0}$ and allows us to apply the same analytic method to our perturbative wet-channel dam break as both problems are locally similar, although in our case the absence of an initial background river flow means that we get a second, almost symmetrical, ``bore'' propagating in the opposite direction that is suppressed in the river case (see equation 3.1 in reference \cite{Berry_2018}). More precisely, our dam break system starts from rest and we are therefore obliged to account for both roots of the Bogoliubov dispersion relationship, namely bi-directional wave propagation. The two roots are 
\begin{gather}
    \omega_{\pm}(k)= v(x,t)k\pm c(x,t)k\sqrt{1+\frac{\xi(x,t)^{2}k^{2}}{4}} \ , \label{eq:bogDisp}
\end{gather}
where the $v(x,t)k$ term accounts for a Doppler shift since we work in a laboratory frame of reference centered at the location of the initial dam, $x=0$. To write down this space and time varying version of the dispersion relation we have assumed a local density approximation so that $c(x,t)=\sqrt{g n(x,t)/m}$ and $\xi(x,t)=\hbar/\sqrt{gn(x,t)m}$ are the local speed of sound and healing length, respectively, and should vary slowly in space and time compared to the wavelength and frequency of the waves. However, as far as the dispersion relation is concerned, for a perturbative dam break the step to $n_{1}$ is a small perturbation on top of $n_{0}$ and the DWs are even smaller in amplitude, so we put $n(x,t)\approx n_{0}$ $\forall (x,t)$, and similarly $c(x,t)\approx c_{0}=\sqrt{g n_{0}/m}$ and $\xi(x,t)\approx\xi_{0}=\hbar/\sqrt{g n_{0} m}$ (we could just as well set the density to be $n_{1}$ $\forall\;(x,t)$, which would result in the same physics). For our choices of parameters $c_{0}\approx 1.8$ mm/s and $\xi_{0}\approx0.8\;\mu$m. 

Similarly, we can approximate $v(x,t)$ in the dispersion relation to its initial value $v(x,t)\approx v_{0}=0$. Applying these approximations to equation (\ref{eq:bogDisp}) we obtain
\begin{gather}
    \omega_{\pm}(k) \approx \pm \sqrt{c_{0}^{2}k^{2}+\frac{c_{0}^{2} \xi_{0}^{2}k^{4}}{4}} \ ,  \label{eq:bogDispScal}
\end{gather}
which is the the basic Bogoliubov dispersion relation given by equation (\ref{eq:bogDispFluid}) for a homogeneous system at rest with the parameters corresponding to those of the lower reservoir. As we shall see, $\omega_{+}$ and $\omega_{-}$ correspond to the frequencies of right and left travelling DWs, respectively. Due to the analogy to relativistic systems, the dam break can be interpreted as a superposition of particles that travel to the right and antiparticles (holes) that travel to the left. 

The dispersion relation for the excitations from the condensate can be canonically quantized to obtain the differential equation obeyed by either the density or the phase fluctuations. Since the dam break problem is posed in terms of initial conditions involving the density, we choose this variable. Squaring both sides of (\ref{eq:bogDispScal}) and making the replacements $\omega_{\pm}\rightarrow \hat{\omega}=i\partial_{t}$ and $k\rightarrow \hat{k}=-i\partial_{x}$, and then applying the resulting operator to the density fluctuation field $\delta n(x,t)$, we obtain a linear partial differential equation for dispersive waves that is second order in $t$ and fourth order in $x$ \cite{Coutant_2012,Coutant_2014,Farrell_2023,Barcelo_2026}
\begin{gather}
    \partial_{t}^{2}\delta n(x,t)=c_{0}^{2}\bigg(\partial_{x}^{2}-\frac{\xi_{0}^{2}}{4}\partial_{x}^{4}\bigg)\delta n(x,t) \ \label{eq:waveEqn}.
\end{gather}
In Appendix \ref{sec:appendixA} we give a more careful derivation based on the quantum hydrodynamic equations. 

\begin{figure}[!t]
	\centering
	\captionsetup{width=1\linewidth}
     \begin{minipage}{0.3\textwidth}
    		\includegraphics[width=6cm]{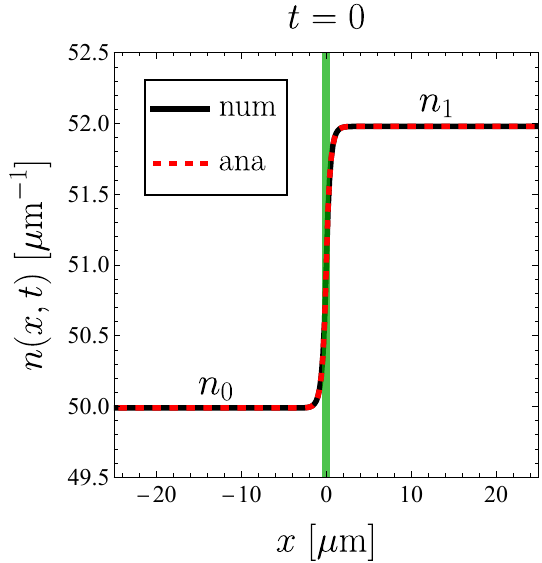} 
    	\end{minipage}%
    	\begin{minipage}{0.3\textwidth}
    		\includegraphics[width=6cm]{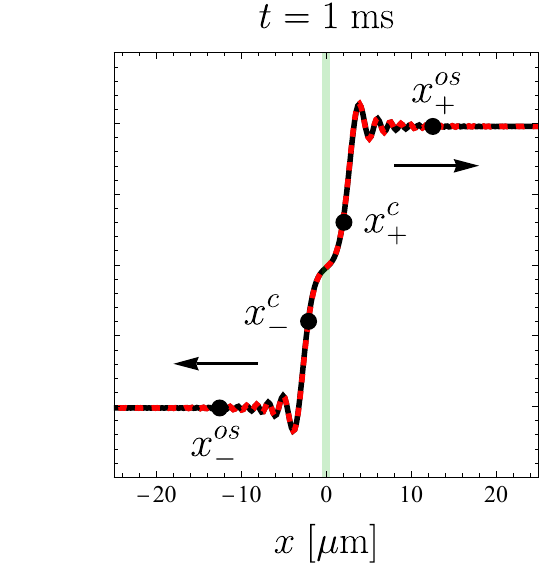} 
    	\end{minipage}%
            \begin{minipage}{0.38\textwidth}
    		\includegraphics[width=6cm]{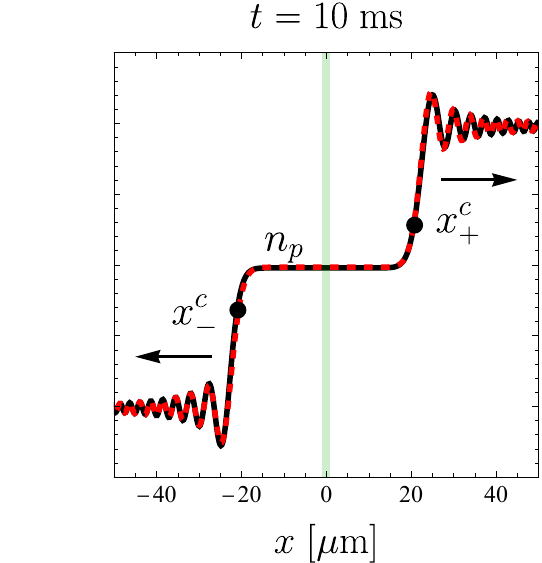} 
    	\end{minipage} \\[0.2cm] 
        \begin{minipage}{0.3\textwidth}
    		\includegraphics[width=6cm]{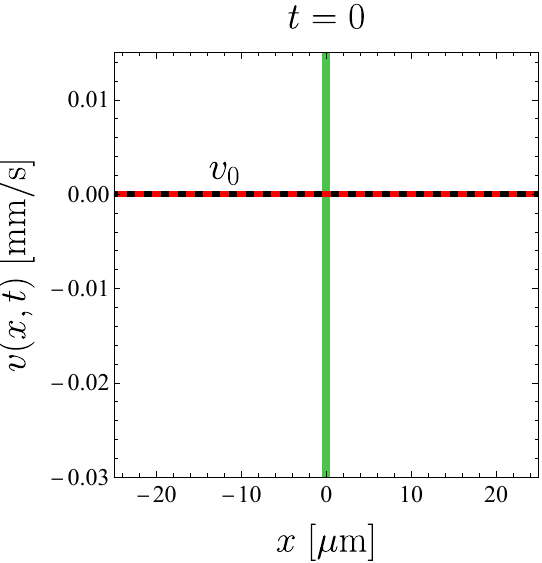} 
    	\end{minipage}%
    	\begin{minipage}{0.3\textwidth}
    		\includegraphics[width=6cm]{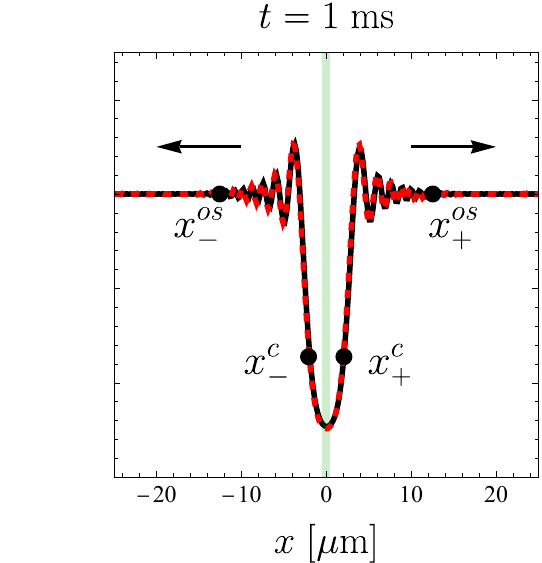} 
    	\end{minipage}%
            \begin{minipage}{0.38\textwidth}
    		\includegraphics[width=6cm]{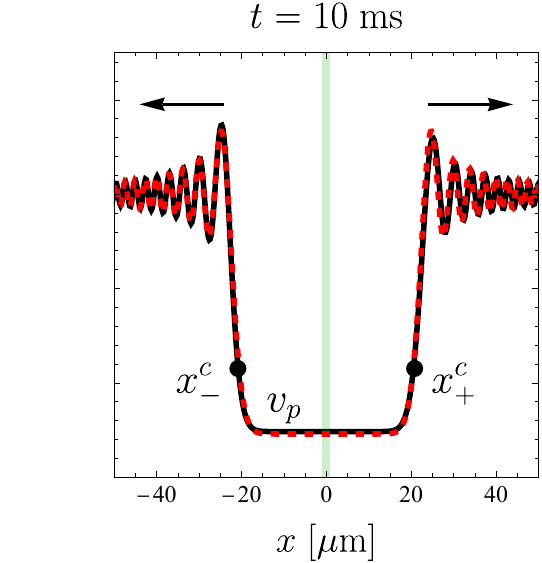} 
    	\end{minipage}\\[-0.2cm]
    \caption{Snapshots of the development of undulatory bore-like waves in the density profile (top row) and flow velocity (bottom) following a perturbative dam break in a BEC, with $n_{0}=50\mu$m$^{-1}$, $\Delta n\approx 0.04 n_{0}\ll n_{0}$, and $\kappa=1/4$. \textbf{Left:} Initial condition at $t = 0$, where the vertical green line represents the position of the initial potential barrier (dam) at $x=0$. \textbf{Middle and Right:} $t=2$ ms and $t=10$ ms after dam breaking. The numerical (black) results are in near exact agreement with the analytic (red dashed) expressions given in equations (\ref{eq:dens}-\ref{eq:etaFunc}) and (\ref{eq:flow}-\ref{eq:deltVb}). The slight asymmetry that can be seen in the numerical curves, particularly in the $t=10$ ms flow plot, vanishes as the initial condition $\Delta n \rightarrow 0$. The points labelled $x_{\pm}^{c}$ and $x_{\pm}^{os}$ correspond to the different wavefronts shown in figure \ref{fig:spaceTime}.} \label{fig:gpeDens}
\end{figure}

 We now construct the solutions to equation (\ref{eq:waveEqn}) in terms of linear superpositions of DWs by writing the total density in the form \cite{Berry_2019}
\begin{gather}
    n(x,t)=n_{0}+\delta n(x,t)=n_{0}+\Delta n \frac{\eta_{+}(x,t)+\eta_{-}(x,t)}{2} \ \label{eq:dens},
\end{gather}
in which the fluctuation terms are given by 
\begin{align}
    \eta_{\pm}(x,t)&=\int_{-\infty}^{\infty}\frac{dk}{2\pi i k f(k)}\textrm{e}^{i(kx-\omega_{\pm}(k)t)} \nonumber \\
    &=\frac{1}{2}+\frac{1}{\pi}\int_{0}^{\infty}\frac{dk}{k f(k)}\textrm{sin}\big(kx-\omega_{\pm}(k)t\big) \ ,\label{eq:etaFunc} 
\end{align}
and where the integration contour in the first line of equation (\ref{eq:etaFunc}) passes below the pole at $k=0$; the pole contributes the 1/2 term in the second line.
It is seen from the general form of the functions $\eta_{\pm}(x,t)$ that they satisfy the dispersive wave equation (\ref{eq:waveEqn}), but to specify the solution describing a dam break we must match the boundary conditions
\begin{eqnarray}
    & &  \eta_{\pm}(x\rightarrow\infty,t)  \rightarrow 1  \quad \forall t \\ 
    & & \eta_{\pm}(x\rightarrow-\infty,t)  \rightarrow 0   \quad \forall t 
 \end{eqnarray}
which ensure that $n(x\rightarrow\infty,t)\rightarrow n_{1}$ and $n(x\rightarrow-\infty,t)\rightarrow n_{0}$, as illustrated in figure \ref{fig:damBreakProblems}. This can be achieved through the choice of the function $f(k)$ which should be an even function of $k$, although the functions $\eta_{\pm}(x,t)$ have the notable property that their asymptotic shapes at long times $t \rightarrow \infty$ do not depend on the details of the initial spatial profile, as will be explained in Section \ref{sec:asymptotic}. The most basic case is $f(k)=1$, which reduces $\eta_{\pm}(x,t)$ to Heaviside step functions at $t=0$. The initial density profile is then discontinuous at $x=0$, but other choices such as polynomials $f(k)=\sum_{j}\alpha_{j}k^{2j}$, allow one to specify different initial density profiles since $\eta_{\pm}(x,0)$ reduces to the Fourier transform of $1/ikf(k)$. In this work we take $f(k)$ to be 
\begin{equation}
    f(k)=1+\kappa \xi_{0}^{2}k^{2} \ , \label{eq:fFunc}
\end{equation} 
which depends on the single dimensionless parameter $\kappa$. If $\kappa$ is real and positive the discontinuity at the step is smoothed, and empirically we find that $\kappa=1/4$ yields an excellent match between the analytic and numerical results at $t=0$, as can be seen in figure \ref{fig:gpeDens}. Thus, both the amplitudes and phases [via equation (\ref{eq:bogDispScal})] of the waves have the same factor of $\xi_{0}^2/4$, in accord with the dispersive term in the differential equation (\ref{eq:waveEqn}).  Substituting equations (\ref{eq:bogDispScal}) and (\ref{eq:etaFunc}) into (\ref{eq:dens}) yields our analytic approximation for $n(x,t)$, which we plot in figure \ref{fig:gpeDens} against the numerical GPE result. There is near exact agreement for all $x$ and $t$.

Although we took the flow velocity to be $v=v_{0}=0$ in the dispersion relation, we can find a perturbative expression for $v(x,t)$ via the coupled equations (\ref{eq:hydro4a}) and (\ref{eq:hydro4b}) given in Appendix \ref{sec:appendixA}. Similarly to equation (\ref{eq:dens}), the flow can be linearized to take the form 
\begin{gather}
    v(x,t)=v_{0}+\delta v(x,t)=v_{0}+\frac{v_{-}(x,t)+v_{+}(x,t)}{2} \ \label{eq:flow},
\end{gather}
where $v_{\pm}(x,t)$ are obtained by taking a spatial derivative of equation (\ref{eq:hydro4b}) and using the quantum flow relation $\delta v(x,t)=\frac{\hbar}{m}\partial_{x}\delta\theta(x,t)$. Rearranging, we find
\begin{align}
    v_{\pm}(x,t) &=-\frac{c_{0}\Delta n}{n_{0}}\int_{0}^{t}(\partial_{x}-\frac{\xi_{0}^{2}}{4}\partial_{x}^{3})\eta_{\pm}(x,t') dt' \nonumber \\
    &=-\frac{c_{0}\Delta n}{n_{0}}\int_{0}^{t}\int_{0}^{\infty}\frac{1}{\pi f(k)}\bigg(1+\frac{\xi_{0}^{2}k^{2}}{4}\bigg)\textrm{cos}\big(kx-\omega_{\pm}(k)t'\big) dk\;dt'  \ .\label{eq:deltVb}
\end{align}
From this, the condensate phase $\theta=\frac{m}{\hbar}\int_{0}^{x} v(x',t)dx'$ can in principle be found, and thus an approximate solution to the GPE (\ref{eq:gpe}) obtained in terms of $\delta n(x,t)$ and $\delta v(x,t)$ via equation (\ref{eq:ordParam}) in Appendix \ref{sec:appendixA}. Plots of analytic versus numerical $v(x,t)$ are also shown in figure \ref{fig:gpeDens}, where the agreement is again seen to be excellent in the perturbative regime. Within the central plateau region with density $n_{p}$, halfway between the densities of the upper and lower reservoirs, the corresponding velocity takes the constant value of $v_{p}$.

\subsection{Standard versus smoothed TF limits}
\label{subsec:smoothed_TF}

\begin{figure}[!b]
	\centering
	\begin{minipage}{0.5\textwidth}
		\centering
		\includegraphics[width=7cm]{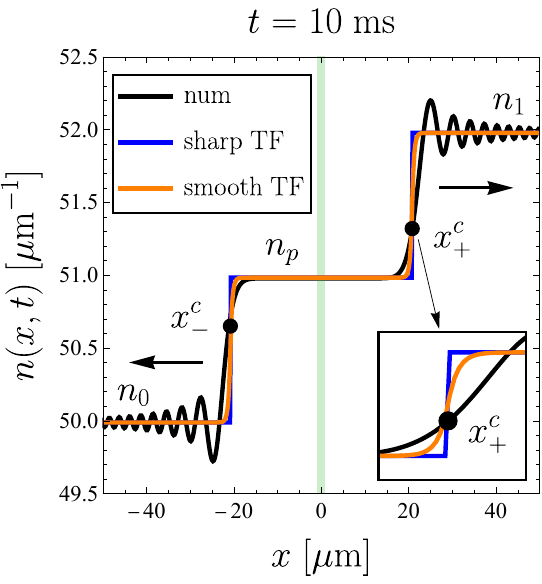}
	\end{minipage}%
	\begin{minipage}{0.5\textwidth}
		\centering
        \includegraphics[width=7.15cm]{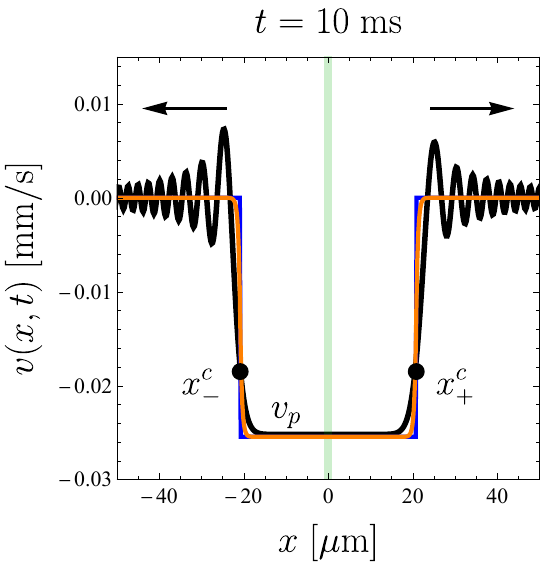}
	\end{minipage}\\[-0.2cm]
	\caption{Comparison of standard and smoothed TF approximations for the perturbative wet-channel dam break. \textbf{Left:} Exact numerical (black) versus both discontinuous sharp (blue) and smoothed (orange) analytic TF limit densities at $t=10$ ms. The inset highlights the differences near the wavefront $x_{+}^{c}$. \textbf{Right:} Same as the left, but for the flow velocity.}\label{fig:TFplot}
\end{figure}

Taking the standard TF limit $\xi\rightarrow 0$ of our dispersive analytic model, equations (\ref{eq:etaFunc}) and (\ref{eq:deltVb}) for the perturbative density profile and velocity simplify to
\begin{align}
    \eta_{\pm}(x,t)&\approx \frac{1}{2}+\frac{1}{2}\textrm{Sgn}\big[(x\mp c_{0}t)/\xi_{0}\big] \ \label{etaFuncTF}, \\
    v_{\pm}(x,t)&\approx -\frac{c_{0}\Delta n}{n_{0}}\frac{\textrm{Sgn}\big[(c_{0}t\mp x)/\xi_{0}\big]\pm \textrm{Sgn}(x/\xi_{0})}{2} \ \label{eq:deltVbTF},
\end{align}
where the \textit{sign} function $\textrm{Sgn}(X)$ equals -1, 0 or 1 when $X<0$, $X=0$, or $X>0$, respectively. A snapshot of these solutions at $t=10$ ms is plotted in figure \ref{fig:TFplot}.  They are discontinuous at the DW fronts $x_{\pm}^{c}(t)=\pm c_{0}t$ for all times. Since our dispersive analytic model only holds in the perturbative regime, its TF limit is also only valid in same regime, but we note that the TF approximation can also be applied to the full continuity and quantum Euler equations (\ref{eq:cont}) and (\ref{eq:euler}) to obtain expressions for $n$ and $v$ for non-perturbative dam breaks \cite{El_2016,Olshanii_2021,Olshanii_2022}. We discuss this solution in Appendix \ref{sec:appendixB}.

If one instead takes the TF limit only within the phase of equation (\ref{eq:etaFunc}) and not the amplitude, we obtain \textit{smoothed} versions of solutions (\ref{etaFuncTF}) and (\ref{eq:deltVbTF}) where the discontinuity has been removed
\begin{align}
    \eta_{\pm}(x,t)&\approx \frac{1}{2}+\frac{1}{2}\bigg(1-\textrm{e}^{-\kappa^{-1/2}\vert (x\mp c_{0}t)/\xi_{0}\vert}\bigg)\textrm{Sgn}\big[(x\mp c_{0}t)/\xi_{0}\big] \ \label{etaFuncTF2}, \\
    v_{\pm}(x,t)&\approx -\frac{c_{0}\Delta n}{2n_{0}}\bigg\{\Big(1-\textrm{e}^{-\kappa^{-1/2}\vert (c_{0}t\mp x)/\xi_{0}\vert}\Big)\textrm{Sgn}\big[(c_{0}t\mp x)/\xi_{0}\big]\pm\Big(1-\textrm{e}^{-\kappa^{-1/2}\vert x/\xi_{0}\vert}\Big)\textrm{Sgn}(x/\xi_{0})\bigg\} \ \label{eq:deltVbTF2} .
\end{align}
Plots of these expressions are included in figure \ref{fig:TFplot}. They give an excellent match to the exact density profile at $t=0$, describing an expanding step which is smooth for all $t$ and also does not contain DWs. The smoothed TF solutions will be useful in Section \ref{sec:rainbows} where we isolate the undulations from the background step.

The greater significance of the TF solutions described in this section is that they confirm that dispersion is responsible for the undulations. This is already apparent from the sound cone in figure \ref{fig:spaceTime} where only DWs are fast enough to outrun the edge of the cone. One can view the TF solutions as describing a BEC expanding \textit{adiabatically} without creating DW excitations. Since dispersion in a BEC is quantum in origin, the undulatory waves are quantum effects.

\section{Asymptotic long-time limit}
\label{sec:asymptotic}

\subsection{Integral-Airy function solutions}

\begin{figure}[!t]
	\centering
	\begin{minipage}{0.5\textwidth}
		\centering
		\includegraphics[width=7.4cm]{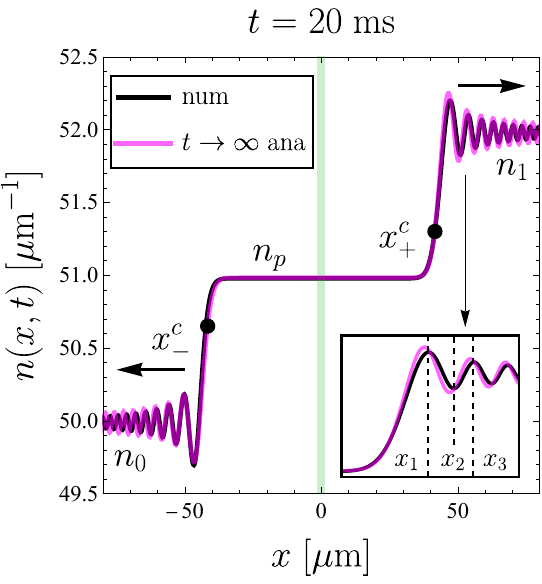}
	\end{minipage}%
	\begin{minipage}{0.5\textwidth}
		\centering
        \includegraphics[width=7.55cm]{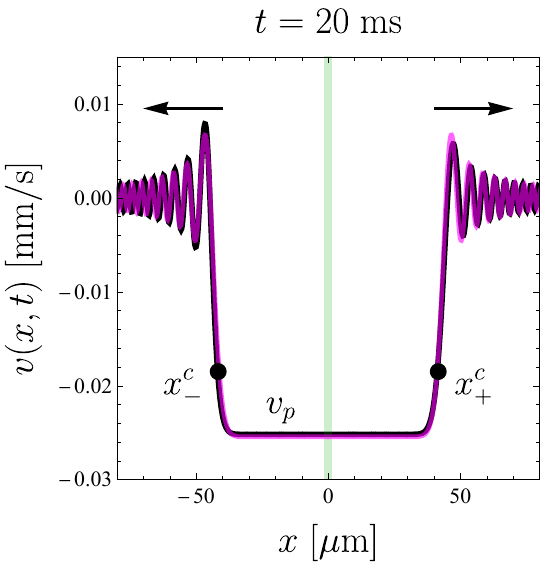}
	\end{minipage}\\[-0.2cm]
	\caption{Integral-Airy long time  asymptotic solution.  \textbf{Left:} Integral of the Airy function (magenta) density via equations (\ref{eq:dens}) and (\ref{eq:airyInt}) versus the numerical solution of the GPE (black), at $t=20$ ms after the dam break. Within the inset we label the positions of the first two crests for the upper DW as $x_{1,3}$, and the first trough by $x_{2}$. \textbf{Right:} Same as the left, but for the flow velocity.}\label{fig:airyPlot}
\end{figure}

The $t \rightarrow \infty$ regime of the undulatory river bore has been discussed by Berry  \cite{Berry_2018,Berry_2019}, where it was shown that the solutions reduce to the integrals of Airy functions. We now apply the same analysis to find the asymptotic long time limit of our perturbative expressions for $n(x,t)$ and $v(x,t)$ while still preserving the effects of dispersion. In particular, as $t\rightarrow\infty$ the factor $\exp[-i\omega_{\pm}(k) t]$ in the integrands of equations (\ref{eq:etaFunc}) and (\ref{eq:deltVb})  will oscillate rapidly causing the integral over $k$ to vanish unless $\omega_{\pm}(k)$ is small. In other words, large wavenumbers (short wavelengths) are most relevant for short times just after dam breaking and play a lesser role during the remainder of the expansion (it is for this reason that at long times the density forgets the details of the initial profile). Thus, in the long time regime we can Taylor expand equation (\ref{eq:bogDispScal}) to third order in $k$
\begin{gather}
    \omega_{\pm}(k)\approx \pm c_{0}k\pm \frac{c_{0}\xi_{0}^{2}k^{3}}{8} \ \label{eq:approxDisp}.
\end{gather}
The cubic term must be retained to preserve quantum dispersive effects, and since the detailed form of $f(k)$ does not matter as $t\rightarrow\infty$, we make the simplest choice $f(k)=1$. Equation (\ref{eq:etaFunc}) then becomes
\begin{gather}
    \eta_{\pm}(x,t) \sim \int_{-\infty}^{\infty}\frac{dk}{2\pi i k}\textrm{e}^{i\big(kx\mp c_{0}kt\mp \xi_{0}^{2}c_{0}k^{3}t/8\big)} \ . \label{eq:eta2}
\end{gather}
Transforming variables $k\rightarrow \big(\frac{8}{3\xi_{0}^{2}c_{0}t}\big)^{1/3}s$, the integral can be expressed as
\begin{align}
    \eta_{\pm}(x,t) = \int_{-\infty}^{\infty}\frac{ds}{2\pi i s}\textrm{e}^{i\left(\zeta_{\pm}(x,t)s\mp \frac{s^{3}}{3}\right)} \ , \label{eq:eta3}
\end{align}
where 
\begin{equation}
    \zeta_{\pm}(x,t)=\frac{2(x\mp c_{0}t)}{(3 \xi_{0}^{2}c_{0}t)^{1/3}} \ .
    \label{eq:zetadefn}
\end{equation}
 As will be shown in more detail shortly, the argument $\zeta_{\pm}(x,t)$ of the $\eta_{\pm}$ functions describes travelling waves of constant speed $c_{0}$ whose wavelengths expand as $t^{1/3}$. Furthermore, given that the integral definition of the Airy function is  $\textrm{Ai}(X)=(1/2 \pi)\int_{-\infty}^{\infty} d\sigma \exp[i(\sigma^{3}/3+\sigma X)]$ \cite{DLMF}, the perturbative solution in equation (\ref{eq:eta3}) can be recognized as the integral of an Airy function
\begin{align}
    \eta_{\pm}(x,t)=\int_{-\zeta_{\pm}(x,t)}^{\infty}\textrm{Ai}(\pm u)du \label{eq:airyInt} \ .
\end{align} 
The resulting prediction for the particle density is plotted against the exact numerical solution of the GPE in figure \ref{fig:airyPlot} for the time $t=20$ ms. We see good agreement, especially near the DW fronts $x_{\pm}^{c}(t)$, even for such a modestly sized value of $t$ after the dam break; the agreement increases for all $x$ as both $t$ increases and for smaller values of $\Delta n$. 

The integral-Airy solution correctly predicts the density $n_{p}$ on the central plateau that forms after a dam break. This can be checked by setting $x=0$, which gives $\zeta_{\pm}(0,t)=\mp 2(c_{0}t/\sqrt{3}\xi_{0})^{2/3}$, and in the limit $t\rightarrow\infty$, the lower bound of equation (\ref{eq:airyInt}) thus tends to $-\zeta_{\pm}(0,t\rightarrow\infty)\rightarrow\pm\infty$. In this case, combining equations (\ref{eq:dens}) and (\ref{eq:airyInt}) simplifies $n(0,t\rightarrow\infty)=n_{p}$ to
\begin{eqnarray}
n(0,t\rightarrow\infty) &  \sim & n_{0}+\frac{\Delta n}{2}\int_{-\infty}^{\infty}\textrm{Ai}(u)du+\frac{\Delta n}{2}\int_{-\infty}^{\infty}\textrm{Ai}(-u)du \nonumber \\
& & =n_{0}+\frac{\Delta n}{2}\int_{-\infty}^{\infty}\textrm{Ai}(-u)du=\frac{n_{0}+n_{1}}{2}  \ , 
\end{eqnarray}
where in the last step we used the fact that $\int_{-\infty}^{\infty} \mathrm{Ai}(u)du=1$. This result for the density on the plateau is as expected: $n_{p}$ is halfway between $n_{0}$ and $n_{1}$. It does not of course hold for non-perturbative dam breaks, see Appendix \ref{sec:appendixB}.

 A characteristic feature of $\eta_{\pm}(x,t)$ as seen in (\ref{eq:eta3}) is its cubic phase function 
 \begin{equation}
 \phi_{\pm}(x,t,s)=\zeta_{\pm}(x,t)s \mp s^3/3  \label{eq:cubicphase} \ . 
 \end{equation}
 The classical rays associated with the wave function can be found from the stationary points of the phase $\partial \phi / \partial s=0$ and hence a cubic phase describes the interference of two rays. In the case of $\eta_{+}$ they occur at
\begin{equation}
s_{1,2}=\pm \sqrt{\zeta(x,t)}=\pm \sqrt{\frac{2(x-c_{0}t)}{(3 \xi_{0}^2 c_{0} t)^{1/3}}}  \ .
\label{eq:rays}
\end{equation}
These two rays coalesce when $x=c_{0}t$, which results in a \textit{fold caustic} and is also the DW wavefront. There is only an evanescent wave with no interference when $x<c_{0}t$.  A similar argument holds for $\eta_{-}$ where the caustic/wavefront occurs at  $x=-c_{0}t$.

\begin{figure}[!b]
    \centering
    \captionsetup{width=1\linewidth}
    \includegraphics[width=12cm]{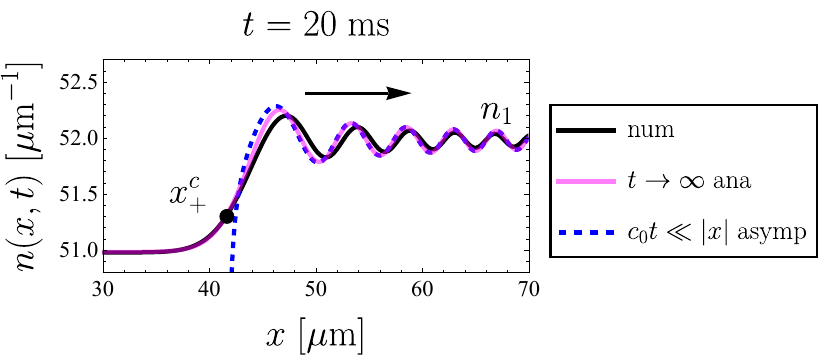}\\[-0.2cm]
    \caption{Comparison between the numerical GPE (black), integral-Airy (magenta), and far-from-wavefront asymptotic (blue dashed) solutions for the upper DW density. The latter solution is obtained by combining equations (\ref{eq:dens}) and (\ref{eq:steep_descent}).} \label{fig:asymp_plot}
\end{figure}

Far outside the caustics, where  $\vert x \vert \gg \pm c_{0}t$,  one can obtain a simple WKB-type approximation to the integral-Airy function solution as \cite{Berry_2019,Kamchatnov_2021}
\begin{equation}
    \eta_{\pm}(x,t)\sim-\frac{1}{\sqrt{\pi}}\big(\pm\zeta_{\pm}(x,t)\big)^{-3/4}\mathrm{cos}\Bigg[\frac{2}{3}\big(\pm\zeta_{\pm}(x,t)\big)^{3/2}+\frac{\pi}{4}\Bigg] \ .
    \label{eq:steep_descent}
\end{equation}
This is plotted as the  dashed blue line in figure \ref{fig:asymp_plot}. We see that it  works well except close to the caustic at $\zeta_{\pm}=0$ where, as usual for a WKB approximation, it suffers an amplitude blow-up. According to the theory of wave catastrophes \cite{Berry_1981,Nye_1999}, the Airy function is the universal wave function that smoothly dresses such fold caustics where two rays coalesce: the interference captured by the cubic phase function regulates the classical singularity giving rise to the Airy fringes, which in our case are the quantum undulations, and in the case of rainbows (see Section \ref{sec:rainbows}) are known as supernumerary arcs \cite{Khare_1974,Nussenzveig_1977,Berry_2015}. The fold caustic is one of the structurally stable catastrophes classified by catastrophe theory and hence occurs generically without the need for tuning \cite{Berry_1981,Gilmore_1981,Nye_1999}. In the present case this means that even if we had expanded the dispersion relation $\omega_{\pm}$ in equation (\ref{eq:approxDisp}) to higher order in $k$, there exists a smooth change of coordinates that maps the phase function near the fold to a cubic polynomial.

The integral-Airy function is also known to provide solutions to the linear version of the Korteweg-de Vries (KdV) equation, which is first order in time and third order in space.  In Appendix \ref{sec:appendixkdv} we discuss how the cubic dispersion relation (\ref{eq:approxDisp}) can be used to map the large $t$ dam break problem discussed here to the linear KdV equation.

\subsection{Scaling properties of quantum undulations}

Another notable feature of the solution (\ref{eq:airyInt}) for $\eta_{\pm}(x,t)$ is that it is a function purely of the single dimensionless quantity $\zeta_{\pm}(x,t)$ in (\ref{eq:zetadefn}). It therefore has the property of being self-similar if we follow the special combination of space and time specified by $\zeta_{\pm}(x,t)=$ constant. Furthermore, in comparison to the simple power laws that are often found in self-similar scaling problems such as phase transitions \cite{Huang_1987}, the integral-Airy function is a non-trivial function of $\zeta_{\pm}$ with oscillatory fringes of varying height and wavelength as well as an exponentially decaying piece.

Let us use these features to obtain 
the long-time scaling properties of the wavelengths $\lambda_{\pm}$ and amplitudes $A_{\pm}$ of the quantum undulations. For brevity we will only focus on deriving $\lambda_{+}$ and $A_{+}$ for the upper DW, but as long as $\Delta n \ll n_{0}$ then $\lambda\approx\lambda_{\pm}$ and $A\approx A_{\pm}$. Because the wavelength changes with position and time, we define $\lambda$ to be the distance between the first two crests closest to the DW front, $\lambda=x_{3}-x_{1}$, where the points $x_{1,3}$ are maxima of $\eta_{+}(x,t)$ as shown in figure \ref{fig:airyPlot}. These stationary points can be obtained by solving $\partial_{x}\eta_{+}(x,t)=0$ for $x$, which gives
\begin{gather}
    \textrm{Ai}\big[-\zeta_{+}(x,t)\big]=0 \ \label{eq:derivCond}.
\end{gather}
Being a special function, the zeros of the Airy function are tabulated \cite{DLMF}. The first three we denote by $z_{1}=-2.338$, $z_{2}=-4.088$, and $z_{3}=-5.521$, each respectively corresponding to points $x_{1,2,3}$ via $z_{1,2,3}=-\zeta_{+}(x_{1,2,3},t)$. The distance between the first two crests of the Airy function is thus $\zeta_{+}(x_{3},t)-\zeta_{+}(x_{1},t)=z_{3}-z_{1}=\Delta z=3.182$, and with this we find the asymptotic long-time expression for the wavelength to be
\begin{gather}
    \lambda(t)=\frac{3^{1/3}\Delta z}{2}(\xi_{0}^{2}c_{0}t)^{1/3}\label{eq:lamb} \ .
\end{gather}
The linear dependence of the healing length $\xi_{0}$ on $\hbar$ implies that $\lambda \propto \hbar^{2/3}$, and thus the wavelength vanishes in the classical limit. Another feature of $\lambda$ is that it grows sub-diffusively as $t^{1/3}$ (i.e.\ slower than the diffusive growth $\sim t^{1/2}$ seen in disordered systems, for example). 

\begin{figure}[!t]
	\centering
	\begin{minipage}{0.5\textwidth}
		\centering
		\includegraphics[width=7.5cm]{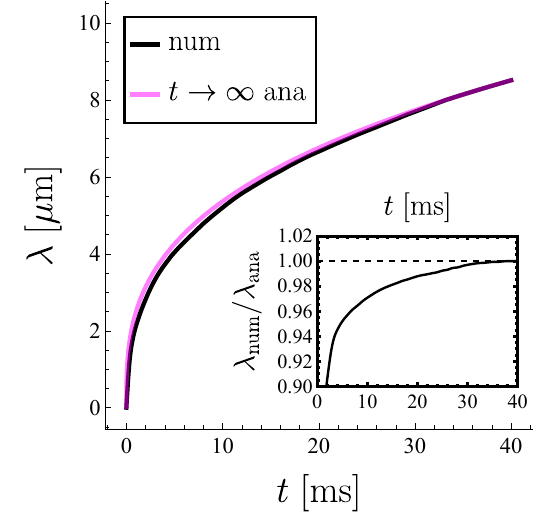}
	\end{minipage}%
	\begin{minipage}{0.5\textwidth}
		\centering
		\includegraphics[width=7.5cm]{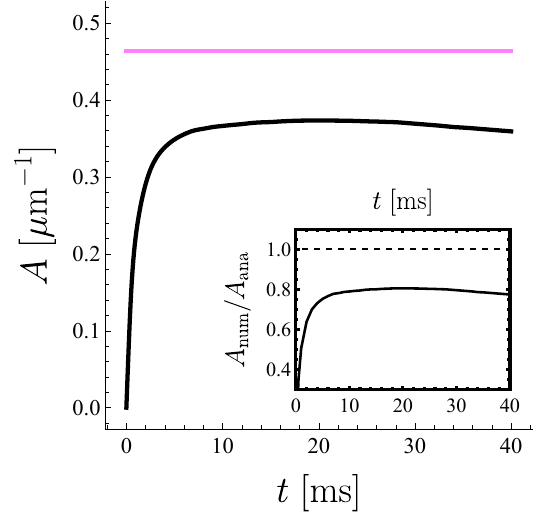}
	\end{minipage}\\[-0.2cm]	
    \caption{Wavelength and amplitude of the quantum undulations as a function of time. \textbf{Left:} Numerical (black) versus long-time analytic (magenta) plots of wavelength $\lambda$ for the perturbative wet-channel dam break. \textbf{Right:} Same as the left, but for amplitude $A$. The insets plot the fractional error between numerical and analytic theories.}\label{fig:numComp1}
\end{figure}

For the amplitude of the undulations, we take the difference in density between the first crest and the first trough, $A=n(x_{1},t)-n(x_{2},t)$. The first and second zeros of the Airy function combined with condition (\ref{eq:derivCond}) yield $\zeta_{+}(x_{1},t)=-z_{1}=2.338$ and $\zeta_{+}(x_{2},t)=-z_{2}=4.088$, and plugging these into equations (\ref{eq:dens}) and (\ref{eq:airyInt}) gives
\begin{gather}
    A=\frac{\Delta n}{2} \Delta\eta_{+} \ \label{eq:amp},
\end{gather}
where $\Delta\eta_{+}=\eta_{+}(x_{1},t)-\eta_{+}(x_{2},t)=0.4661$. This is \textit{constant} in time and is simply set by the initial magnitude of the dam break $\Delta n$. In 2D or 3D we expect the wave height to decay with time as it spreads outwards, but in 1D the height is maintained for each point on the wave, i.e.\ for each fixed value of $\zeta_{\pm}(x,t)$.  

In figure \ref{fig:numComp1} we plot the analytic expressions given in equations (\ref{eq:lamb}) and (\ref{eq:amp}) against their numerical counterparts for increasing $t$. Of course, they are both expected to disagree at short times because the analytic results are limited to the long time solution. Nevertheless, we see $\lambda$ is well described at all times whereas there are significant discrepancies for $A$. In this regard, it is worth remembering that we retained the $k$-cubic dispersion within the phase in the integral-Airy profile (\ref{eq:airyInt}), but entirely neglected it in the amplitude. Although $A$ does improve initially as $t$ increases, the agreement eventually begins to slightly decrease. For smaller values of $\Delta n$ the agreement between numerical and long-time analytic values of both $\lambda$ and $A$ increases overall.

Finally, to connect back to the undulatory tidal bore phenomenon in rivers, we see from equations 5.3 and 5.4 in reference \cite{Berry_2019} that the wavelength of the water undulations behind the bore front also grows as $t^{1/3}$. However, the scaling of the amplitude and wavelength with the natural length scale provided by the river depth $d_{r}$ is different to our dependence on $\xi_{0}$. This is due to the difference between the shallow-water dispersion relation given in equation
(\ref{eq:waterdispersion}) and the Bogoliubov dispersion relation.  A $t^{1/3}$ scaling also occurs in the expansion of spin-chain domain walls \cite{Fagotti_2017,Bulchandani_2018,Bulchandani_2019,Kirkby_2019,Wei_2022,Riddell_2023,McRoberts_2024}, and classical tsunami propagation \cite{Jeffreys_1999}. Indeed, $t^{1/3}$ scaling universally arises in perturbative dynamics with step-like initial conditions, whether for classical or quantum waves. Its origin can be traced back to the cubic phase function given in equation (\ref{eq:cubicphase}) and thus we can see that the reason for this universality is the structural stability of fold caustics.

\subsection{Asymptotics of the flow velocity}

To obtain an asymptotic long-time approximation for the flow velocity, we neglect the cubic derivative term within the first line of equation (\ref{eq:deltVb}), giving the perturbative flow as $\delta v(x,t)\approx -\frac{c_{0}^{2}}{n_{0}}\int_{0}^{t}\partial_{x}\delta n(x,t')dt'$. This is equivalent to the well known Josephson relationship between density and phase fluctuations in superfluids \cite{Pethick_2008}, and can be seen by integrating both sides of the above approximate expression with respect to $x$ and using the definition $\frac{\hbar}{m}\delta \theta(x,t)=\int \delta v(x,t) dx$ to find $\delta \theta(x,t)\approx -\frac{g}{\hbar}\int_{0}^{t}\delta n(x,t')dt' \implies \partial_{t}\delta\theta(x,t)\approx -g\delta n(x,t)/\hbar$ which for our system implies $ \partial_{t}\theta(x,t)\approx -gn(x,t)/\hbar$, precisely the Josephson relationship. Plugging equation (\ref{eq:airyInt}) into equation (\ref{eq:deltVb}) and neglecting the cubic derivative thus produces 
\begin{equation}
    v_{\pm}(x,t) \sim -\frac{c_{0}\Delta n}{n_{0}}\int_{0}^{t}\frac{2\textrm{Ai}\big(\mp\zeta_{\pm}(x,t')\big)}{3^{1/3}(\xi_{0}^{2}c_{0}t')^{1/3}} dt' \ \label{eq:deltVairy}.
\end{equation}
The spatial dependence of this flow pattern is plotted against the exact numerical solution of the GPE in figure \ref{fig:airyPlot}. Equation (\ref{eq:deltVairy}) can be used to find the flow on the central plateau as $v_{p}=v(0,t\rightarrow\infty)\sim -c_{0}\Delta n/(2n_{0})$. 

\section{Non-perturbative dam breaks and dispersive wave amplification}
\label{sec:nonperturbative}

\begin{figure}[!t]
	\centering
	\begin{minipage}{0.475\textwidth}
		\centering
		\includegraphics[width=6.95cm]{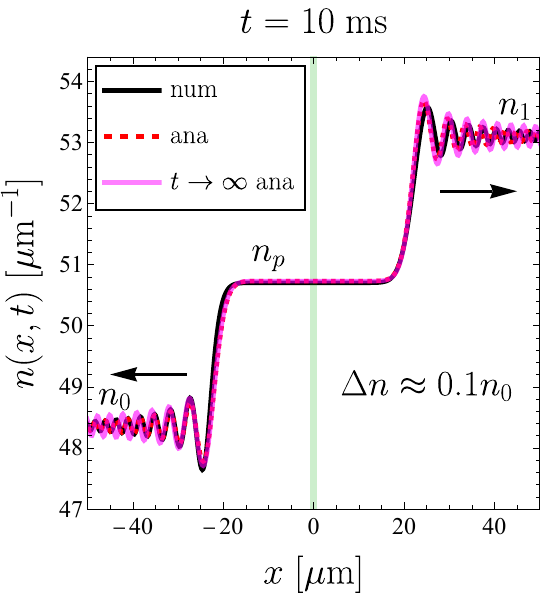}
	\end{minipage}%
	\begin{minipage}{0.475\textwidth}
		\centering
		\includegraphics[width=6.95cm]{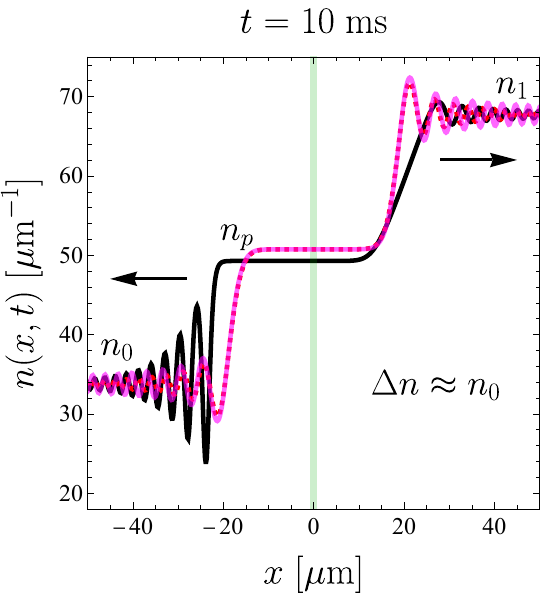}
	\end{minipage}\\
 \centering
	\begin{minipage}{0.475\textwidth}
		\centering
		\includegraphics[width=7.45cm]{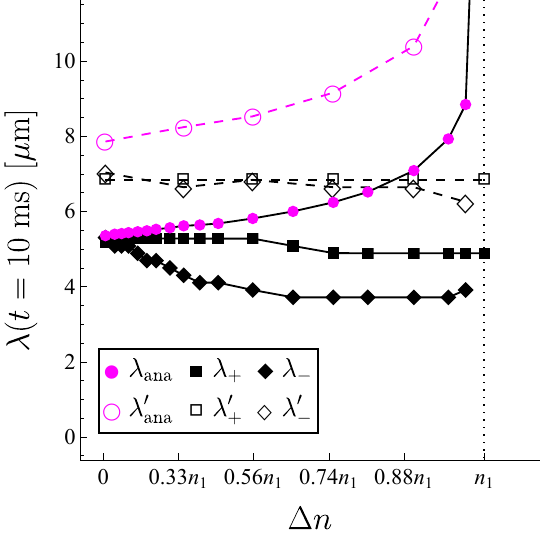}
	\end{minipage}%
	\begin{minipage}{0.475\textwidth}
		\centering
        \includegraphics[width=7.45cm]{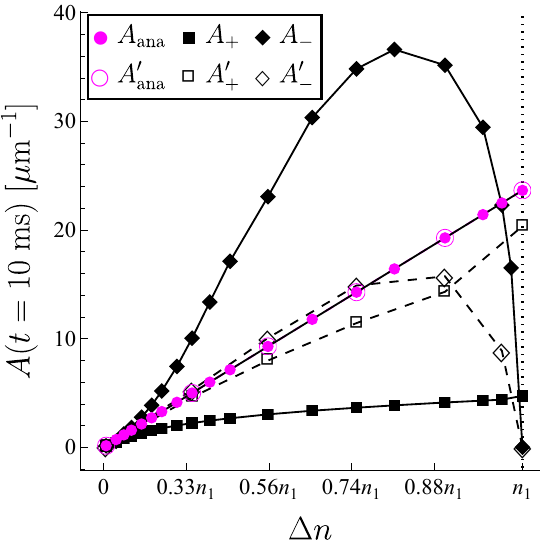}
	\end{minipage}\\[-0.2cm]
	\caption{Perturbative versus non-perturbative dam breaks. \textbf{Top row:} Density snapshots at $t=10$ ms, comparing numerical simulation (black) with both arbitrary-$t$ (red dashed) and long-time (magenta) perturbative analytic theories, for $\Delta n \approx0.1 n_{0}$ (left) and $\Delta n \approx n_{0}$ (right). \textbf{Bottom row:} Undulation wavelength $\lambda$ (left) and amplitude $A$ (right) as a function of $\Delta n$ (does not increase linearly as a fraction of $n_{1}$) at $t=10$ ms. The perturbative analytic predictions are plotted in magenta and the exact numerical results [capturing the different behaviour of the upper ($+$, squares) and lower ($-$, diamonds) waves] are plotted in black. The leftmost points correspond to a very perturbative dam break with $\Delta n\approx 0.01 n_{0}\approx 0.01n_{1}$, while the rightmost points (vertical dotted lines) denote the largest possible value of $\Delta n\approx n_{1}$: a dry-channel dam break. The hollow points, representing the primed variables $\lambda'$ and $A'$, describe the case where the scattering length has been quenched from its initial value $a(t\leq 0)=100 a_{0}$ to $a(t>0)=10 a_{0}$ at the time of dam breaking in order to enhance the undulations.}\label{fig:non_pert_plots}
\end{figure}

The condensate parameters we use in this paper are realistic for modern day experiments \cite{Roati_2007,Duchene_2026}. However, small perturbations such as $\Delta n\approx 0.04 \ll n_{0}$ would be challenging to observe in comparison to background noise and fluctuations. It is therefore also important to consider non-perturbative dam breaks. In the top row of figure \ref{fig:non_pert_plots} we plot snapshots at fixed time of numerical versus analytic densities for the two cases $\Delta n\approx0.1 n_{0}$ and $\Delta n\approx n_{0}$. In the former case the analytic theory still does a fairly good job, while the latter is beyond the perturbative regime. Similar plots can be obtained for the flow but we omit them henceforth for brevity. 

It is clear that as $\Delta n$ increases, both the perturbative arbitrary-$t$ (\ref{eq:etaFunc}) and asymptotic long-time (\ref{eq:airyInt}) theories begin to disagree with the numerical GPE results. In the perturbative regime the upper and lower DWs can be treated independently because $\delta n$ obeys a linear equation [equation (\ref{eq:waveEqn}) or equivalently equation (\ref{eq:KdV}) in the $t\rightarrow\infty$ limit]. In the non-perturbative regime the symmetry between upper and lower waves is broken and the dynamics become non-linear as the lower DW transitions into a DSW. The TF solution also breaks down by becoming multi-valued in the region where the DSW forms \cite{El_2016}. This occurs because the higher speed of sound in the plateau region (due to higher density) relative to the lower region causes the left edge of the plateau to overtake the adjoining edge of the lower region leading to a breaking wave. The upper DW does not suffer from this instability. 

In the bottom row of figure \ref{fig:non_pert_plots} we compare the perturbative predictions versus the exact (numerical) results for the wavelengths $\lambda_{\pm}$ and amplitudes $A_{\pm}$ of the upper and lower undulations for increasing values of $\Delta n$ (which does not increase linearly as a fraction of $n_{1}$) at fixed time $t=10$ ms. For our choice of parameters one can safely apply the perturbative $t\rightarrow\infty$ analytic theory up to an initial density difference of approximately $\Delta n\approx0.25n_{0}\approx0.2n_{1}$. Even past this, our analytic expression (\ref{eq:lamb}) for the wavelength still yields a correct order of magnitude for a generous range of non-perturbative $\Delta n$, particularly for the upper DW. As $\Delta n$ approaches the most non-perturbative situation possible, namely the dry-channel regime $\Delta n\approx n_{1}$, the disagreement becomes substantial since the lower DSW vanishes (see also figure \ref{fig:damBreakProblems}) while the analytic wavelength diverges. For more details regarding the wet- to dry-channel transition, see figure 15 (and the surrounding text) in reference \cite{Kamchatnov_2021}.

At the time of writing, the experimental observation of DWs and DSWs in freely expanding condensates has not yet been achieved to the best of our knowledge. The most favourable situation for visibility would be to set $\Delta n$ so as to hit the peak shown in the lower right panel of figure \ref{fig:non_pert_plots} of the amplitude $A_{-}$ of the lower wave (DSW). Waves at this peak result from a non-perturbative and yet wet-channel value of $\Delta n$. Two opposing factors affect the amplitude of the undulations: 
\begin{enumerate}
\item Going deeper into the TF regime (smaller $\xi$) implies smaller $A$
\item Having a steeper initial density step $\Delta n/\Delta x$ generates larger $A$.
\end{enumerate}
These two effects are connected since the maximum steepness of the density step is set by $\xi$, which occurs in the case of a dam potential which is a step function (infinitely steep). Empirically, we find that in situations where the dam potential is a step function (which is the case for all the figures shown in this paper) the TF regime `wins' in the sense that $A$ gets smaller as $\xi$ decreases (at least for non-perturbative dam breaks). However, if a softer dam potential is used at fixed $\xi$ then $A$ diminishes. 

A relatively simple experimental trick that can be used to generate larger DWs for a given $\Delta n$ is to initialize the scattering length $a$ to a large value prior to breaking, and subsequently quench it to a lower value precisely at the time of dam breaking. This both increases the initial dam steepness, and then moves the system more into the quantum dispersive regime once evolution begins. The result of this procedure on $\lambda$ and $A$ is also shown in the bottom row of figure \ref{fig:non_pert_plots} for $a(t\leq 0)=100 \, a_{0}$ before breaking and $a(t>0)=10 \, a_{0}$ afterwards. The wavelength increases to a roughly constant value for both upper and lower waves, until near where the transition from wet- to dry-channel initial conditions occurs, $\Delta n \approx0.9n_{1}\approx9n_{0}$. The analytic approximation for $\lambda$ is significantly worse under this protocol, but the analytic result for $A$ matches both numerical results for both the upper and lower waves well, up until $\Delta n \approx 0.75 n_{1}\approx3n_{0}$. Significantly, the upper wave is amplified in size while the lower wave is suppressed, but only for non-perturbative values of $\Delta n$. In the non-perturbative regime DW amplification and DSW suppression is expected since the upper wave is purely quantum dispersive, while the lower balances dispersion with non-linearity. 

\section{Rainbows and black hole analogues}
\label{sec:rainbows}

The Airy-type wave functions we have found thus far are usually associated with underlying singular ray-caustic structures, so it is worth studying the rays emitted by the dam break. Caustic formation in BECs has previously been studied in a variety of theoretical and experimental works \cite{Rooijakkers_2003,Chalker_2009,Huckans_2009,ODell_2012,Rosenblum_2014,Goldberg_2019,Mossman_2021,Agarwal_2023}, but a novel feature of the caustics discussed in this paper is that their resolution relies on the dispersive effects caused by quantum pressure. Caustics also provide a novel way to understand event horizons since they correspond to bifurcations where rays coalesce/are born.

\subsection{Logarithmic Airy function}

Before describing the caustic structure associated with the integral-Airy wave functions, it is useful to compare them against a close relative, the \textit{logarithmic-Airy function} which has been derived in the context of Hawking radiation with dispersion \cite{Coutant_2012,Coutant_2014,Farrell_2023}. The integral-Airy wave function does not include the effects of the condensate flow velocity on the dispersion relation as these are small in the perturbative regime. However, in non-perturbative dam breaks these corrections can be significant, a prime example being when the flow velocity exceeds the speed of sound such that a sonic event horizon is formed, as will be discussed in detail in Section \ref{sec:blackhole}. Logarithmic-Airy functions describe event horizons and do take fluid flow into account. They are defined via the integral
\begin{equation}
\psi_{LA}(x,\omega)=\int_{-\infty}^{\infty} \frac{dk}{2 \pi k} \mathrm{e}^{i[k^3d^3/3+kx- (\omega/\kappa_{H}) \log (k d) ]}  \ ,
\label{eq:logairy}
\end{equation}
see equation (12) of \cite{Farrell_2023}. Here, $\omega$ is the frequency in the laboratory frame, and is related to the frequency $\omega'$ in the fluid rest frame through the Galilean-boosted dispersion relation $\omega'=\omega-v(x)k$. Equation ($\ref{eq:logairy}$) is obtained by expanding the dispersion relation to third order, just as we did in equation (\ref{eq:approxDisp}), and also assumes that the fluid flow velocity $v(x)$ varies linearly about a horizon located at $x=0$, so that $v(x)=-c+\kappa_{H} x$, where $\kappa_{H}$ is the velocity gradient (which is analogous to the surface gravity in astrophysical black holes \cite{Barcelo_2026}) and should not be confused with the smoothing parameter $\kappa$ used in equation (\ref{eq:etaFunc}). The length scale $d=(c \xi^2 /8 \kappa_{H})^{1/3}$ corresponds to a finite width (broadening) of the horizon which arises due to the effects of dispersion. 

$\psi_{LA}(x,\omega)$ gets its name because of the extra logarithmic term in the phase in comparison to the Airy function, and which comes about physically from a Doppler shift of the wavenumber $ \propto 1/x$ as waves approach the horizon. We therefore see a family resemblance between the integral-Airy wave function $\eta_{\pm}(x,t)$ in equations (\ref{eq:etaFunc}) and (\ref{eq:eta3}) and the logarithmic-Airy function: $\eta_{\pm}(x,t)$ corresponds to the zero-frequency ($\omega=0$) case such that the logarithmic term does not appear (which is connected to the fact that there is no horizon in $\eta_{\pm}$), the choice $f(k)=1$ is made, and also the coordinate $x$ in the phase of $\psi_{LA}(x,\omega)$ is replaced by $\zeta_{\pm}(x,t)$. On the other hand, $\eta_{\pm}(x,t)$ has time dependence whereas $\psi_{LA}(x,\omega)$ is constrained to be time independent by the assumption that the flow $v(x)$ is stationary. Both assume the background condensate density is spatially constant. The differences and similarities can also be appreciated by examining the differential equation obeyed by the logarithmic-Airy function $(i \omega/\kappa_{H} - x \partial_{x}+d^3 \partial_{x}^{3} ) \psi_{LA} =0$, which is a modified version of the linear KdV equation discussed in Appendix \ref{sec:appendixkdv}.

\subsection{Caustic structure of dispersive waves: rainbows}

\begin{figure}[!t]
	\centering
    \captionsetup{width=1\linewidth}
    \includegraphics[width=13cm]{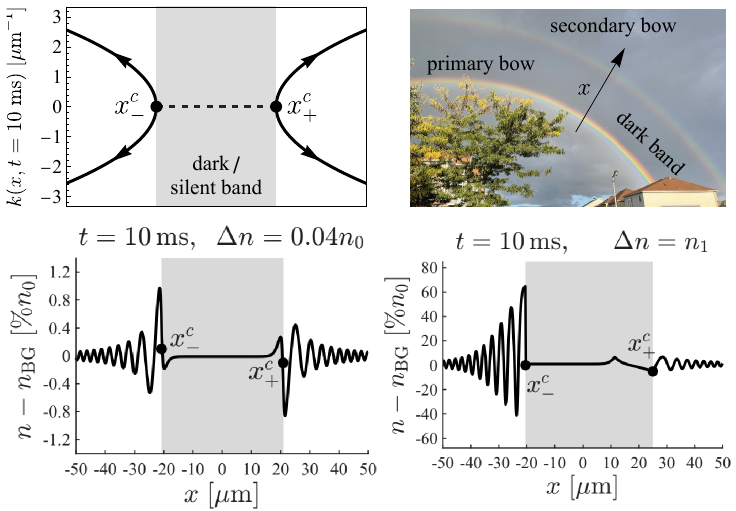}\\[-0.2cm]
    \caption{Double rainbow.  \textbf{Top left:} Classical rays plotted in phase space ($x,k$) for a perturbative dam break at fixed time $t=10$ ms. Two rays pass through each position $x$ (except in the dark/silent band) producing interference fringes (undulations) in the wave function (\textbf{bottom left}). The wavefronts at $x_{\pm}^{c}$ also correspond to two oppositely facing fold-caustics where rays coalesce. One complex ray enters the dark/silent band from each caustic (dashed line), giving rise to two exponentially small evanescent waves. \textbf{Top right:} A perpendicular cross section through an optical double rainbow gives the same qualitative ray structure, where each rainbow corresponds to a caustic with Alexander's dark band sandwiched between. \textbf{Bottom right:} A non-perturbative dam break leads to undulations of unequal magnitude on each side which more closely resembles the asymmetric structure of a double rainbow but for different reasons. Note that in the bottom row the background TF density steps have been subtracted off so only the undulations are shown.}
    \label{fig:doubleRainbow}
\end{figure} 

The caustic structure of the logarithmic-Airy function has been extensively examined in reference \cite{Farrell_2023}, including for $\omega=0$ where it coincides with the integral-Airy function. In this case the classical rays give rise to a fold caustic, but as shown in figure 3 of \cite{Farrell_2023} the wave profile is slightly different from a standard Airy function and takes the form of a step function dressed by Airy fringes via the pole in the amplitude and the two saddle points in the phase, respectively. Thus, each DW of the perturbative dam break corresponds to an Airy function modulated by a step, and each DW front $x_{\pm}^{c}(t)=\pm c_{0}t$ to a fold caustic. The latter point can be seen by finding the rays as described in equation (\ref{eq:rays}), which gives 
\begin{equation}
x\mp c_{0}t\mp 3c_{0}\xi_{0}^{2}k^{2}t/8=0 \ .
\label{eq:raysolution}
\end{equation}
A snapshot of these rays is plotted in the top left panel in figure \ref{fig:doubleRainbow} which shows the 2D phase space for rays provided by the coordinates ($x,k$) at the instant of time $t=10$ ms. We see that there are two rays passing through each spatial point $x$ (except in the silent band) with equal and opposite $k$ values. In a semiclassical description it is the interference between these two rays that gives rise to the Airy fringes/quantum undulations. At the two wavefronts $x^{c}_{\pm}$ the rays coalesce giving rise to fold caustics.

The quintessential example of a fold caustic is the optical rainbow, where the Sun's rays are focused by water droplets in the sky \cite{Khare_1974,Nussenzveig_1977,Berry_2015}. In Appendix \ref{sec:appendixrainbow} we briefly outline the geometric optics involved and show that the same characteristic local ray behaviour occurs for either of the two dam break DWs as in a rainbow, but with the role of position ($x$) exchanged for angle in the sky. The main point is that pairs of rays entering a water droplet at different points output at the same angle, except at the special angle $\theta_{\mathrm{primary}}\approx 42^{\circ}$ above the horizontal where the rays concerned coalesce to give a caustic where the intensity peaks.   

\subsection{Double rainbows and silent bands}

The primary bow shown in figure \ref{fig:doubleRainbow} occurs when the Sun's rays enter a water droplet and are internally reflected once before exiting. The secondary bow is formed by another family of rays that undergo two internal reflections before exiting (higher-order rainbows also occur but are increasingly faint). Referring to Appendix \ref{sec:appendixrainbow} for details, the total ray structure including back-to-back fold caustics that we see in the top left panel in figure \ref{fig:doubleRainbow} for the pair of DWs emitted in a perturbative dam break corresponds to the signature ray and caustic structure of a double rainbow as shown in figure \ref{fig:rainbowRays}. In both the optical and DW cases, no real rays from these two families of rays can exist in the gap between the two caustics, which in the optical case is Alexander's dark band as seen in the photo in the top right of figure \ref{fig:doubleRainbow} (the band is darker than either of the rainbows that surround it). In the BEC case this is a region where no sound waves propagate, i.e.\ a `silent band'. It is precisely the intermediate plateau $n_{p}$, corresponding to the region inside the sonic cone in figure \ref{fig:spaceTime} at a particular instant of time. As time runs on, the silent band becomes wider. Only complex rays can exist in the dark/silent band and these are associated with evanescent waves which are captured by the exponential tails of the Airy functions on each side. 

Remarkably, the same double rainbow ray and wave structure can be used to explain dynamical phase transitions following a sudden quench in a spin system \cite{Link_2024}. In that case, the exponentially small overlap between the initial quantum state and the quantum state at the critical time $t=t_{c}$ after the quench, which is captured by the Loschmidt echo $L(t)=\vert \langle \psi (0) \vert \psi(t) \rangle \vert^2$, gives rise to a `quantum dark band'. The same basic phenomenon also occurs in limit shape phase transitions sometimes referred to as Arctic circles \cite{Pallister_2022} and the emptiness formation probability in large deviation theory \cite{Arzamasovs_2019}.

Dispersion plays different roles in dam breaks and optical rainbows. In rainbows dispersion leads to the angular separation of the colours of light because each wavelength in sunlight undergoes different refraction in a raindrop such that the Airy functions for different colours are superimposed, but with a relative shift. What we predominantly see as the `colours of the rainbow' is just the main peak of each different Airy function. The subdominant fringes for each colour are not visible in the photo in figure \ref{fig:doubleRainbow} but are sometimes observed as the faint supernumerary arcs where the colours repeat. By contrast, the DWs created in a perturbative dam break have a single set of fringes as clearly seen in the bottom row of figure \ref{fig:doubleRainbow}. Even though the wavepackets $\eta_{\pm}(x,t)$ in equation (\ref{eq:etaFunc}) are made from an integral over all wavenumbers, each wavelength only contributes to a single point $x$ at a single moment of time, as shown in the top left panel of figure \ref{fig:doubleRainbow}. Taking vertical cuts through this picture, each spatial point outside of the silent band gets contributions from just two rays $\pm k$ with equal wavelengths. According to the ray equation (\ref{eq:raysolution}), as time progresses the wavefronts/caustics move apart and the parabolas of rays become narrower so that the sound reaching any point $x$ is red-shifted over time.

Note that in both panels in the bottom row of figure \ref{fig:doubleRainbow} we have removed the `TF steps' from the density profiles so that only the undulations are plotted in order that the analogy to the double rainbow be clearer (there are no steps in the optical case). The left hand panel is for the perturbative case and is obtained by taking the numerical GPE density and subtracting off the smoothed TF theory given in Section \ref{subsec:smoothed_TF} by equations (\ref{etaFuncTF2}) and (\ref{eq:deltVbTF2}). The right hand panel is for the non-perturbative case where we do not have a simple prescription for smoothing. We have therefore subtracted the arbitrary-$\Delta n$ TF theory profile of equation (\ref{eq:trueTFA}) from Appendix \ref{sec:appendixB} in order to isolate for the upper DW and lower DSW. The arbitrary-$\Delta n$ TF profile and its derivative possess non-physical discontinuities resulting in two kinks in the density difference near $x_{+}^{c}$ and a sudden discontinuity at $x_{-}^{+}$.

\subsection{Dynamic horizons in dam breaks}
\label{sec:blackhole}

Sonic horizons are closely analogous to black hole event horizons, as first pointed out by Unruh within a general hydrodynamic framework \cite{Unruh_1981,Barcelo_2026}.
If the flow velocity $v(x)$ in a dam break exceeds the local speed of sound $\vert v(x) \vert > c(x)$ then a horizon forms at that point, but as dam breaking problems are time dependent one can have dynamic horizons whose location and/or effective surface gravity changes in time. Dynamic horizons have recently been studied in various analogue systems \cite{Kolobov_2021,Fabbri_2021,Fourdrinoy_2022,Balbinot_2022}, but of particular relevance to the present paper is the experiment by Sharan \textit{et al} on a dry-channel dam break in a quasi-1D BEC \cite{Sharan_2025}. The presence of a shallow harmonic trapping potential along the direction of the long axis in the experiment caused the dam break waves to eventually reflect back, theoretically leading to additional horizons that displayed interesting dynamics, including horizon mergers. 

Here we restrict our analysis to completely free 1D expansion with no harmonic or other potential present. Horizon formation not only occurs with dry-channel dam breaks, but can also happen in wet-channel dam breaks: in reference \cite{Cao_2025} wet-channel horizon formation and Hawking radiation were studied theoretically within a polariton condensate. It was found that a horizon formed within the rarefaction wave (RW) at the location where the initial dam was located, as well as multiple other horizons within the DSW. For simplicity, in the following we do not concern ourselves with the multiple DSW horizons (and in the dry-channel case there is only a single horizon).

\begin{figure}[!t]
	\centering
	\begin{minipage}{0.35\textwidth}
		\centering
		\includegraphics[width=6cm]{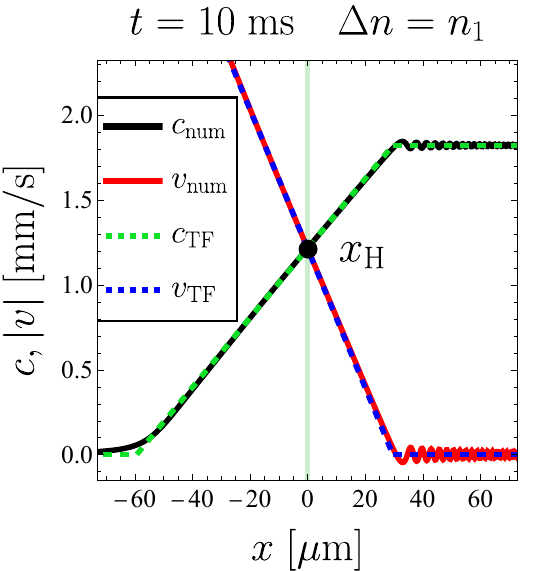}
	\end{minipage}%
	\begin{minipage}{0.35\textwidth}
		\centering
		\includegraphics[width=6cm]{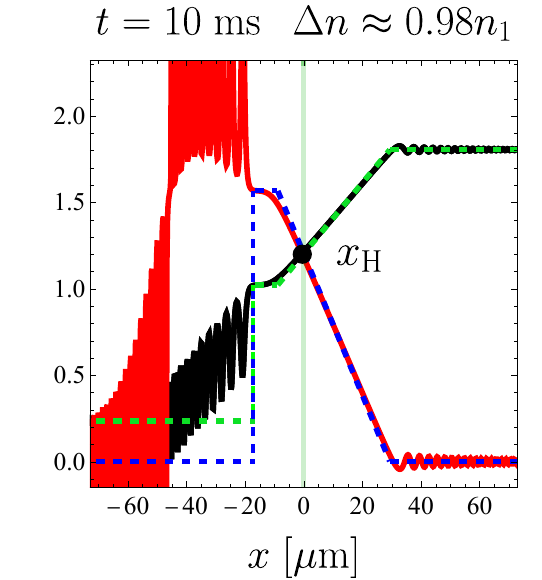}
	\end{minipage}\\[0.2cm]
 \centering
	\begin{minipage}{0.35\textwidth}
		\centering
		\includegraphics[width=6cm]{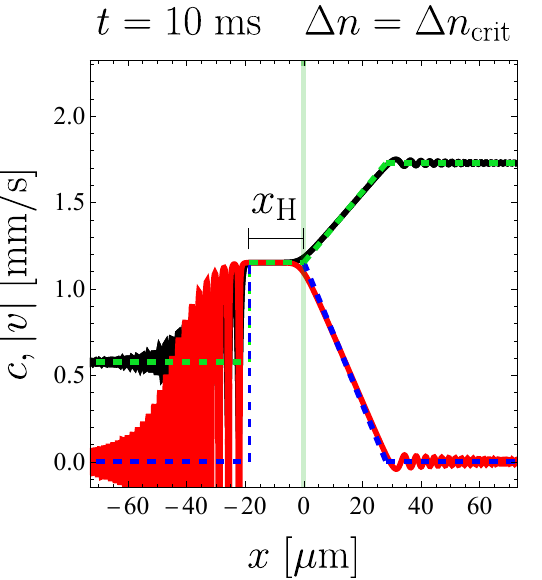}
	\end{minipage}%
	\begin{minipage}{0.35\textwidth}
		\centering
        \includegraphics[width=6cm]{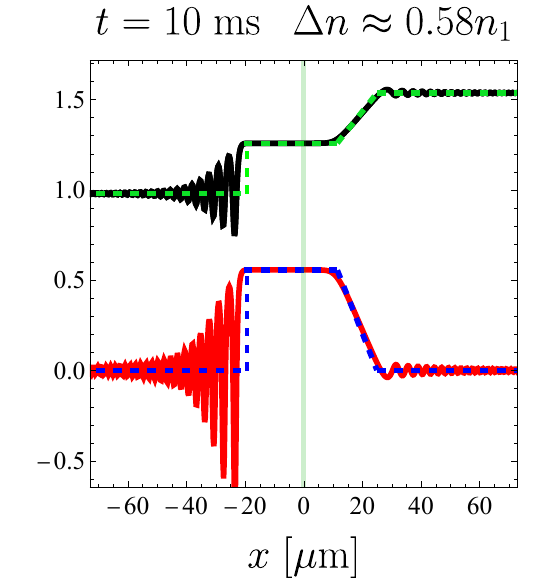}
	\end{minipage}\\[-0.2cm]
    \caption{Snapshots at $t=10$ ms after a dam break of the flow velocity $v(x)$ and speed of sound $c(x)$ spatial profiles. The horizon is at position $x_{H}$ where $c(x)=\vert v(x)\vert $. \textbf{Top left:} Dry-channel. \textbf{Top right:} Wet-channel with horizon. \textbf{Bottom left:} An extended horizon occurs exactly when $\Delta n= \Delta n_{\mathrm{crit}}$, which for our choice of parameters is $\Delta n_{\mathrm{crit}}\approx81\mu$m$^{-1}$. \textbf{Bottom right:} Wet-channel without horizon.} \label{fig:horizons}
\end{figure}

We find that a horizon forms during dam breaking for density differences greater than a critical value $\Delta n_{\mathrm{crit}}$, i.e.\ for both dry- and certain wet-channel scenarios, which we calculate in the TF approximation below. Figure \ref{fig:horizons} plots $c(x)$ and $\vert v(x)\vert$ for different values of $\Delta n$ at a given instant of time, comparing the exact numerical solution of the GPE against the arbitrary-$\Delta n$ TF theory given in Appendix \ref{sec:appendixB}. In figure \ref{fig:horizonsB} we plot the time dependence of the corresponding horizon location $x_{\mathrm{H}}(t)$ for each case. 

In the TF approximation the location of the horizon is fixed at $x_{\mathrm{H}}^{\mathrm{TF}}=0$ $\forall t$, exactly where the initial dam was located before breaking. The horizon lies within the RW part of the TF solution where the flow speed depends linearly on position [as was assumed to be the case in order to construct the logarithmic-Airy function in equation (\ref{eq:logairy})] and can be expressed as
\begin{gather}
    v(x,t)=-c_{\mathrm{H}}+\kappa_{\mathrm{H}}(t) x , \  \label{eq:linFlow}
\end{gather} 
where $\kappa_{\mathrm{H}}(t)=\partial_{x}v\vert_{x=0}=2/3t$ is the time dependent effective surface gravity (flow velocity gradient) of the horizon, not to be confused with the smoothing parameter $\kappa$ from before, and $c_{\mathrm{H}}=c(0,t)=2c_{1}/3$ is the speed of sound at the horizon. $\kappa_{\mathrm{H}}(t)\propto t^{-1}$ decreases in time and is therefore somewhat analogous to the surface gravity of a gravitational Schwarzschild black hole that has a mass $M$ which is increasing over time since $\kappa_{\mathrm{grav}}(M)\propto M^{-1}$. The time-reversed situation corresponds to a black hole that is decreasing in mass (i.e. radiating). 

\begin{figure}[!b]
    \centering
    \captionsetup{width=1\linewidth}
    \includegraphics[width=11cm]{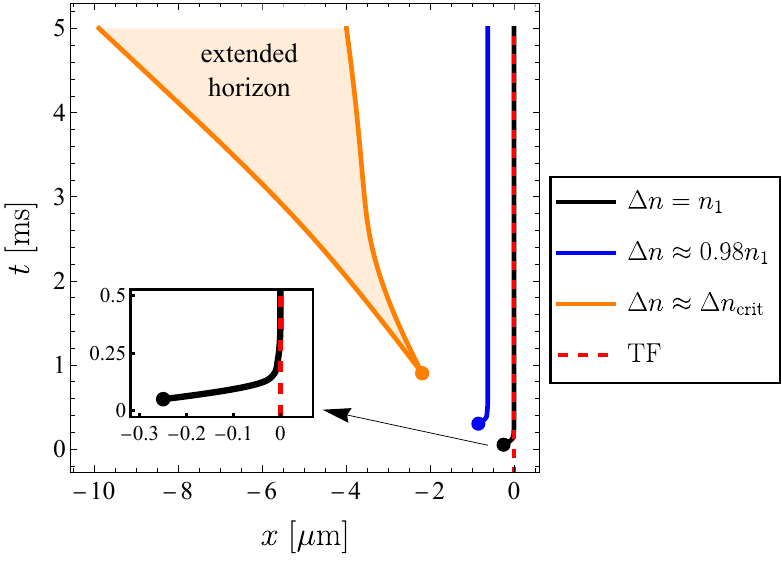}\\[-0.2cm]
    \caption{Plots of the location of the horizon  $x_{H}(t)$ after dam breaking found by solving the GPE numerically for the scenarios in figure \ref{fig:horizons} that possess a horizon. The dots indicate the initial time and location where the horizon forms. For $\Delta n=\Delta n_{\mathrm{crit}}$ an extended horizon forms and expands over time, with the shaded region showing the internal part. The inset highlights the dry-channel case  and shows where and how the horizon moves at early times, before settling to the constant TF value.} \label{fig:horizonsB}
\end{figure}

Numerical solutions of the GPE show that for a brief period of time after dam breaking no horizon exists for any value of $\Delta n$, see figure \ref{fig:horizonsB}. When a horizon does form, it is always to the left (downstream) of the initial dam location $x=0$, and quickly settles to a constant value as time progresses. The timescale over which the horizon settles is quantum in origin since it is not present in the TF theory, but is accounted for by the GPE \eqref{eq:gpe}. Its order of magnitude can be estimated from the energy-time Heisenberg uncertainty relationship, $\Delta E\Delta t\sim\hbar/2$. Taking the characteristic spatial scale of the density variation to be $\xi$ gives $\partial_x^2\sim\xi^{-2}$ in the quantum-pressure term $\propto\hbar^{2}$ of equation \eqref{eq:euler}, yielding the associated energy scale $\Delta E\sim\hbar^2/(2m\xi^2)=m c^{2}/2$ and subsequently $t_{\mathrm{formation}} \approx \Delta t  \sim\hbar/(m c^{2})$, analogous to the Compton timescale, i.e.\ the time required for light to travel over one Compton wavelength. For the parameters used in our dry-channel dam break we find $\Delta t\approx0.5$ ms, which is of the same order as the timescale seen in the inset of figure \ref{fig:horizonsB}.

In the dry-channel case $x_{\mathrm{H}}^{\mathrm{num}}(t\rightarrow \infty)\rightarrow x_{\mathrm{H}}^{\mathrm{TF}}=0$; the long time limit is precisely the regime where the TF theory is accurate (thus, any deviation from $x_{\mathrm{H}}=0$ can be considered a quantum effect here). Strikingly, exactly at $\Delta n=\Delta n_{\mathrm{crit}}$ a horizon of finite extent forms within the intermediate plateau/band of silence, refer to the bottom left panel in figure \ref{fig:horizons} and also the shaded region of figure \ref{fig:horizonsB}. This horizon extends over a range of $x$ at a given $t$ as opposed to the other cases where the horizon is only located at a single point in space (in 1D). If a wave were to be externally created within the extended horizon in the opposite direction to the flow it would presumably remain largely frozen in place, unable to move. Extended horizons have recently been studied in reference \cite{Penalver_2025,Penalver_2026}. 

In the TF regime one can derive the condition for horizon formation for an arbitrary dam break by setting $c_{p}=v_{p}$ (see Appendix \ref{sec:appendixB}), finding that 
\begin{equation}
    \Delta n \geq \Delta n_{\mathrm{crit}} = 8n_{0}=\frac{8}{9} n_{1} , \label{eq:horizon_cond}
\end{equation}
or equivalently $n_{1}\geq9n_{0}$. At $t=0$ the chemical potential $\mu$ must be constant $\forall(x,t)$, so that the value in the upper and lower regions must be equal, $\mu_{0}=\mu_{1}$. In the TF regime this implies $g n_{0}+V_{0}=g n_{1}+V_{1}$ and lets us find in general that $g\Delta n=-\Delta V$, and therefore for a horizon to form the condition $g n_{0}\geq-\Delta V/8$ or equivalently $g n_{1}\geq-9\Delta V/8$.

It is interesting to note that tidal bores have a flow structure analogous to white holes rather than black holes \cite{Berry_2018,Volovik_2005,Volovik_2006}. Referring to figure 2 in reference \cite{Berry_2018}, the height profile of the bore is basically the same as the density profile in the top left panel of our figure \ref{fig:horizons}, and the inside of the `hole' is also on the left hand side of the horizon since that is where the flow speed is supercritical ($\vert v \vert > \sqrt{g_{\mathrm{grav}}d_{r}}$). However, due to the background river flow, the direction of the flow is opposite to the dam break case. This means that rather than being sucked into the black hole, waves are instead swept out of it.

\section{Breaks without dams: imprinted flow scheme}
\label{sec:imprintedflow}

\begin{figure}[!b]
	\centering
    \includegraphics[width=15.5cm]{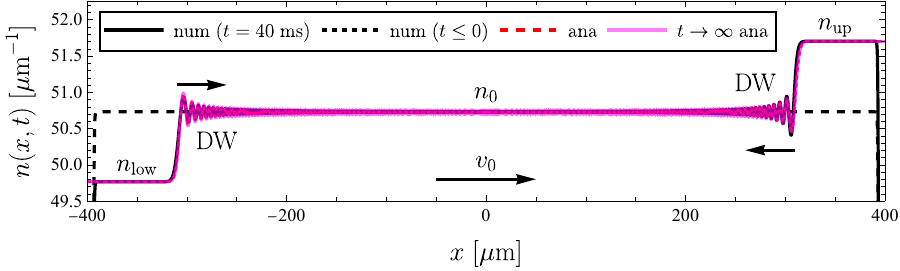}\\
    \includegraphics[width=15.5cm]{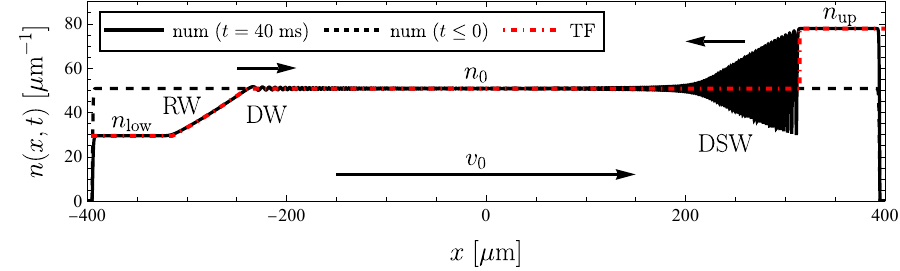}\\[-0.2cm]
	\caption{Densities plotted at $t=40$ ms after a flow $v_{0}$ is imprinted on a BEC in a box trap. \textbf{Top:} Perturbative case with $v_{0}\approx0.02 c_{0}$ gives $n_{\mathrm{up}}-n_{\mathrm{low}}\approx 0.04 n_{\mathrm{low}}\ll n_{0}$. We compare numerical results (black) against analytic arbitrary-time theory (red dashed),  and long-time theory (magenta).  Note that the density at the edges drops to zero, albeit outside of the plotting range. \textbf{Bottom:} Non-perturbative case with $v_{0}\approx0.31 c_{0}$ and $n_{\mathrm{up}}-n_{\mathrm{low}}\approx 1.65 n_{\mathrm{low}}$. We compare numerical results (black) against  TF theory (red dot-dashed, via Appendix \ref{sec:appendixB}). The edge density goes to zero within the plotting range.}\label{fig:splitEmission}
\end{figure}

We now point out an alternative experimental scenario for creating dam break wave structures without a dam break. Consider a 1D BEC trapped in a box trap of length $L$ \cite{Navon_2021}. This results in a condensate of uniform bulk density $n_{0}$ with edges that smoothly tend to zero over the length scale $\xi_{0}$. At $t=0$ a positive rightwards flow $v_{0}>0$ is suddenly imposed everywhere while the outer trap walls of the box remain in place. This situation was studied in reference \cite{Dubessy_2021} in the context of generalized hydrodynamics, and can be achieved by imprinting a linearly varying phase across the BEC using a laser or magnetic potential gradient which is then switched off. The dynamics following the velocity quench is shown in figure \ref{fig:splitEmission} for two different values of $v_{0}$. Alternatively, the same situation could be obtained from the fluid frame of reference by sweeping both the left and right potential walls leftwards with constant velocity. If the left wall were to instead remain fixed in place so that only the right wall was swept leftwards, only the right DW (or DSW) would form, i.e.\ a quantum mechanical piston \cite{Hoefer_2008,Mossman_2018,Schuttelkopf_2026}.

Examining the figure we see that the motion of the condensate following the quench leads to a reduced density plateau of constant value $n_{\mathrm{low}}$ at the left edge near $-L/2$ that travels to the right, while at the other end, near the right edge at $L/2$, the flow causes a raised density plateau of constant value $n_{\mathrm{up}}$, which travels to the left. These plateaus are accompanied by DWs, DSWs, and RWs, depending on the value of $v_{0}$. Therefore, the combined effect of the initial density steps at the edges $\pm L/2$ and the imposed initial flow $v_{0}$ is such as to generate flow and density patterns like those that occur in a standard wet-channel dam break with initial conditions  $\Delta n>0$ in the middle and $v_{0}=0$. However, it is as though the central plateau has been split in half, with each half moved to opposite ends of the system. The values of $n_{\mathrm{low}}$ and $n_{\mathrm{up}}$ depend on the magnitude of $v_{0}$, which controls whether the ``dam break'' is perturbative or non-perturbative.

For the perturbative situation (upper panel in figure \ref{fig:splitEmission}) we also plot both our arbitrary-time and asymptotic long-time theories and see excellent agreement with the numerical simulation as before, while in the non-perturbative wet-channel situation where the perturbative theory does not match as well, we instead compare with the arbitrary-$\Delta n$ TF theory (see Appendix \ref{sec:appendixB}). 

There are some subtleties that must be accounted for when comparing analytic theory with simulation in figure \ref{fig:splitEmission}: i) Because we have a finite value of $v_{0}>0$ which we previously took to be zero, equations (\ref{eq:etaFunc}), (\ref{eq:airyInt}), and (\ref{eq:trueTFA}) must be modified to include the resulting Doppler shift $v_{0}k$ that we had previously ignored. ii) The density of the analytic solutions must be translated by $\pm L/2$ so that each wave emanates from its respective edge. iii) Due to the fact that the edges of the initial condensate drop to zero density over a finite distance given by the healing length, there is a slight position offset $\Delta x_{\mathrm{edge}}$ of the order of $\xi_{0}$ that must be accounted for. In the case of figure \ref{fig:splitEmission}, $\Delta x_{\mathrm{edge}}\approx 0.007 L$. iv) A final major difference of the constant flow scheme is that there is no a situation completely analogous to a dry-channel dam break; although sufficiently large $v_{0}$ results in $n_{\mathrm{low}}\rightarrow 0$, $n_{\mathrm{up}}$ is always larger than $n_{0}$ and thus comes with a DSW.

\section{Concluding remarks}
\label{sec:conclusion}

Quantum dam breaks are a type of local quench that launches out-of-equilibrium many particle dynamics akin to those studied over the last decade in a range of cold atom and trapped ion experiments \cite{Cheneau_2012,Langen_2013,Jurcevic_2014,Eisert_2015,Cevolani_2018}. We have studied dam breaks in a quasi-1D BEC with an emphasis on the role of dispersion and caustics which represent quantum and classical aspects of the problem, respectively. Using numerical solutions of the Gross-Pitaevskii equation (GPE) we are able to model a range of different physical situations from perturbative to non-perturbative regimes, and verified analytic approximations for the density and flow velocity valid in the perturbative limit in terms of integrals of Airy functions that are motivated by an analogy to undular tidal bores in rivers.

Perturbative dam breaks generate a sound cone in spacetime with edges defined by wavefronts that travel outwards at the speed of sound. The cone has the characteristic feature of being inside-out in the sense that the density profile is completely flat inside (no sound waves) while the undulations lie outside and are due to dispersive waves with a wavelength that grows in time as $(\hbar^2 t)^{1/3}$. The Airy function is known to be the universal wave function that dresses fold caustics where a pair of rays coalesce and this is confirmed here by an analysis of the rays underlying the dispersive waves, the caustics being located at the two opposing wavefronts.

The caustic and wave structure of perturbative dam breaks in BECs is in close analogy to the naturally occurring optical phenomenon of double rainbows: the two bows are back-to-back fold caustics separated by Alexander's dark band that only evanescent light waves can enter. The angular separation of the two bows in the sky maps onto the spatial distance between the two wavefronts in the BEC and the dark band maps to a `silent band' which is the inside of the sonic cone. Furthermore, the Airy functions that dress the rainbows (derived by G.B. Airy for precisely this purpose \cite{Airy_1838}) also have their oscillations outside the dark band, showing that the quantum undulations outside the sonic cone are analogous to the supernumerary arcs of rainbows.   The BEC dam break case has the added feature of being dynamic because the width of the silent band grows in time at twice the speed of sound.

In dry-channel [and wet-channel dam breaks when $\Delta n > \Delta n_{\mathrm{crit}} \sim (8/9)n_{1}$] we found that a dynamic sonic horizon forms a short time after dam breaking and eventually settles near the location of the initial dam. At the critical density difference $\Delta n=\Delta n_{\mathrm{crit}}$ a spatially extended horizon forms within the intermediate plateau. Dam breaks in BECs could offer a way to study analogue gravity that provides a more intricate and realistic analogy to certain gravitational black holes, including dynamics where the surface gravity is increasing or decreasing over time. 2D sonic horizons whose surface gravity periodically transitions in time from decreasing to increasing could also in principle be studied in 2D systems \cite{Olshanii_2021,Olshanii_2022}, as well as related scenarios in polariton condensates \cite{Cao_2025}. In contrast to tidal bores propagating up rivers that have analogous properties to white holes, dam breaks create flows analogous to black hole spacetimes.

Regarding the experimental visibility of the effects discussed in this paper, we found that quenching the scattering length to lower values at the time of dam breaking significantly amplifies the height of the quantum undulations. Also, a quite different experimental scheme for realizing dam break waves was proposed where a flow velocity is suddenly imposed on uniform BEC trapped in a box trap. In this case it is the density steps at the edges combined with the imposed flow that creates dam break-like wave dynamics.

\section*{Acknowledgments}
The authors would like to thank Markus Oberthaler, Jelte Duch\^{e}ne, Michael Berry, Christopher Howls, Michael Forbes, Dimitry Pelinovsky, Ana Mucalica, Felix Kaufmes, Jay Mehta, and John Wolfram for useful discussions. The authors would also like to thank the Natural Sciences and Engineering Research Council of Canada (NSERC; Ref.\ RGPIN-2025-06703) for funding.

\section*{Competing Interests Statement}
The authors have no competing interests to declare.

\appendix

\section{Perturbative dispersive wave equation}
\label{sec:appendixA}

Here we derive the 2nd order in $t$ and 4th order in $x$ linear partial differential equation that describes collective excitations (DWs) on the back of a BEC following a perturbative dam break. We follow the derivation performed in reference \cite{Barcelo_2026}.

We recall that for a BEC the Bogoliubov approximation breaks $\psi(x,t)$ into a large condensed ground state component $\psi_{0}(x,t)$ and a small background perturbation $\delta\psi(x,t)$ via the linearization $\psi(x,t)\approx
\psi_{0}(x,t)+\epsilon\delta\psi(x,t)$, where we introduce the dimensionless parameter $\epsilon$ for book-keeping purposes and later remove it by setting it to unity. Equivalently, in hydrodynamical form this amounts to replacing both $n(x,t)\rightarrow n_{0}(x,t)+\epsilon\delta n(x,t)$ and $\theta(x,t)\rightarrow \theta_{0}(x,t)+\epsilon\delta\theta(x,t)$ in equations (\ref{eq:cont}) and (\ref{eq:euler}), the $\theta$ transformation being equivalent to $v(x,t)\rightarrow v_{0}(x,t)+\epsilon\delta v(x,t)$ after taking a position derivative. Neglecting terms greater than first order in $\epsilon$ yields two sets of coupled partial non-linear differential equations, the first of which comes from collecting the zeroth order $\epsilon^{0}$ terms and is
\begin{gather}
    \partial_{t}n_{0}+\frac{\hbar}{m}\partial_{x}(n_{0}\partial_{x}\theta_{0})=0 \label{eq:hydro2a} \ , \\
    \hbar\partial_{t}\theta_{0}+\bigg(\frac{\hbar^{2}}{2m}(\partial_{x}\theta_{0})^{2}+V+g n_{0}-\frac{\hbar^{2}}{2m\sqrt{n_{0}}}\partial_{x}^{2}\sqrt{n_{0}}\bigg)=0 \ , \label{eq:hydro2b}
\end{gather}
while the second set comes from collecting the first order $\epsilon$ terms and is
\begin{gather}
    \partial_{t}\delta n+\frac{\hbar}{m}\partial_{x}(n_{0}\partial_{x}\delta\theta+\delta n\partial_{x}\theta_{0})=0 \label{eq:hydro3a} \ , \\
    \hbar\partial_{t}\delta\theta+\frac{\hbar^{2}}{m}(\partial_{x}\theta_{0})(\partial_{x}\delta\theta)+g\delta n-\frac{\hbar^{2}}{4m}\hat{\mathcal{D}}\delta n=0 \ , \label{eq:hydro3b} 
\end{gather}
where the operator $\hat{\mathcal{D}}$ is defined by
\begin{gather}
    \hat{\mathcal{D}}\delta n = \frac{\partial_{x}^{2}\delta n}{n_{0}}-\frac{(\partial_{x}n_{0})(\partial_{x}\delta n)}{n_{0}^{2}}-\frac{(\partial_{x}^{2}n_{0})\delta n}{n_{0}^{2}}+\frac{(\partial_{x}n_{0})^{2}\delta n}{n_{0}^{3}} \ \label{eq:extraTerms}.
\end{gather}
Equations (\ref{eq:hydro2a}) and (\ref{eq:hydro2b}) are the same as (\ref{eq:cont}) and (\ref{eq:euler}), but for $n_{0}$ and $\theta_{0}$ instead of $n$ and $\theta$. Linearizing $\psi=\sqrt{n}\textrm{e}^{i\theta}$ to first order in $\epsilon$ (and setting $\epsilon=1$ after) yields an approximate solution of the 1D GPE in terms of the $n_{0}$, $\theta_{0}$, $\delta n$, and $\delta \theta$:
\begin{equation}
    \psi=\sqrt{n}\textrm{e}^{i\theta}\approx \sqrt{n_{0}}\textrm{e}^{i\theta_{0}}+\bigg(\frac{\delta n}{2\sqrt{n_{0}}}+i\sqrt{n_{0}}\delta\theta\bigg)\textrm{e}^{i\theta_{0}} \ \label{eq:ordParam}.
\end{equation}

Equation (\ref{eq:extraTerms}) follows the notation of reference \cite{Barcelo_2026} and is the quantum pressure term. One can combine equations (\ref{eq:hydro3a}) and (\ref{eq:hydro3b}) into a single linear partial differential equation for $\delta\theta$ that is 2nd order in $t$ and 4th order in $x$ by setting temporal and spatial gradients of both the background density and flow velocity to zero. This is a good approximation when $n_{0}$ and $v_{0}$ vary little over $t$ and $x$ compared to how much the perturbations $\delta n$ and $\delta v$ do, which is the case for a perturbative wet-channel dam break. Applying this assumption, equations (\ref{eq:hydro3a}) and (\ref{eq:hydro3b}) simplify to 
\begin{gather}
    (\partial_{t}+v_{0}\partial_{x})\delta n+\frac{n_{0}\hbar}{m}\partial_{x}^{2}\delta\theta=0 \ , \label{eq:hydro4a} \\
    (\partial_{t}+v_{0}\partial_{x})\delta\theta+\frac{g}{\hbar}\bigg(1-\frac{\hbar^{2}}{4 m n_{0}g}\partial_{x}^{2}\bigg)\delta n=0 \ \label{eq:hydro4b}.
\end{gather}
Acting $\frac{n_{0}\hbar}{m}\partial_{x}^{2}$ upon equation (\ref{eq:hydro4a}) and using equation (\ref{eq:hydro4b}) to replace the $\delta\theta$ term with one proportional to $\delta n$, we can alternatively combine the coupled equations into single partial differential equation for $\delta n$ (as opposed to $\delta\theta$):
\begin{gather}
    (\partial_{t}+v_{0}\partial_{x})^{2}\delta n=\frac{n_{0}g}{m}\bigg(\partial_{x}^{2}-\frac{\hbar^{2}}{4m n_{0}g}\partial_{x}^{4}\bigg)\delta n  \  \label{eq:waveEqnAppendix}.
\end{gather}
We are able to additionally set $n_{0}$ as constant and $v_{0}\approx 0$ as discussed in Section \ref{sec:perturbative}, which yields equation (\ref{eq:waveEqn}).

In the case of a non-perturbative dam break like those considered in Section \ref{sec:nonperturbative}, gradients of $n_{0}$ and $v_{0}$ can no longer be neglected in equations (\ref{eq:hydro3a}) and (\ref{eq:hydro3b}), and one cannot easily simplify the problem into a single linear partial differential equation such as (\ref{eq:waveEqnAppendix}). One must instead aim to solve equations (\ref{eq:hydro3a}) and (\ref{eq:hydro3b}), or convert the coupled problem into a single non-linear differential equation which includes the effects non-linearity and dispersion (i.e.\ DSWs), such as a non-linear KdV equation \cite{El_2016,Mohapatra_2026}.

\section{TF solutions for non-perturbative dam breaks}
\label{sec:appendixB}

In Section \ref{sec:perturbative} we obtain the TF limit of our dispersive analytic expressions for a perturbative dam break. However, these are only valid in the perturbative regime when $\Delta n\ll n_{0}$. The TF solutions can also be obtained for arbitrary $\Delta n$ by first taking the $\xi\rightarrow 0$ limit of equations (\ref{eq:euler}). Setting $n=v^{2}/4$ and $v=2u/3$ \cite{El_2016,Olshanii_2021,Olshanii_2022}, these two equations can be brought into the form of a single inviscid Burgers equation
\begin{gather}
    \partial_{t}u+u\partial_{x}u=0 \ \label{eq:burgerEq},
\end{gather}
equivalent to the dispersionless limit of a non-linear KdV equation. Along with other related non-linear partial differential equations (e.g., the GPE itself or the sine-Gordon equation \cite{Agarwal_2023}), equation (\ref{eq:burgerEq}) is often applied to model classical shockwaves via the method of characteristics \cite{Chanson_2004}, and can be modified to include dissipation or dispersion in classical or quantum fluids respectively \cite{Hoefer_2006}. The only known explicit solution of equation (\ref{eq:burgerEq}) is the self similar expression $u(x,t)=x/t$, which for an arbitrary dam break scenario yields the true arbitrary-$\Delta n$ TF theory. These expressions possess a region that is multi-valued due non-linear wave-breaking: where the true TF theory breaks down and heralds the formation of a DSW \cite{El_2016}. For our purposes we can replace the multi-valued part of the arbitrary-$\Delta n$ TF theory by a step at the shock front (located at $-c_{p} t$), yielding expressions
\begin{equation}
  n(x,t) \approx \begin{cases} 
          n_{1}, & c_{1}t<x \\
          \frac{m}{9g}\Big(\frac{x+2 c_{1}t}{t}\Big)^{2}, & (3-\sqrt{n_{1}/n_{0}})c_{0}t/2<x<c_{1}t \\
          n_{p}, & -c_{p}t<x<  (3-\sqrt{n_{1}/n_{0}})c_{0}t/2 \\
          n_{0},  & x<-c_{p}t\label{eq:trueTFA} \
       \end{cases},
\end{equation}
\begin{equation}
  v(x,t) \approx \begin{cases} 
          0, & c_{1}t<x \\
          \frac{2}{3}\Big(\frac{x-c_{1}t}{t}\Big), & (3-\sqrt{n_{1}/n_{0}})c_{0}t/2<x<c_{1}t \\
          v_{p}, & -c_{p}t<x<  (3-\sqrt{n_{1}/n_{0}})c_{0}t/2  \\
          0,  & x<-c_{p}t \ \label{eq:trueTFB} 
       \end{cases}.
\end{equation}
The moving boundaries of each component of the above piecewise equations can be found via the method of characteristics, as described in the context of classical hydrodynamic dam breaks \cite{Chanson_2004} or via the method of Whitham modulation theory \cite{Kamchatnov_2021}, see the latter reference's figure 13. For arbitrary dam break reservoir difference $\Delta n$, the plateau density and flow are given respectively by $n_{p}=(\sqrt{n_{0}}+\sqrt{n_{1}})^{2}/4$ and $v_{p}=c_{0}-c_{1}$, see equation 216 in reference \cite{Kamchatnov_2021}. The third and fourth rows of equations (\ref{eq:trueTFA}) and (\ref{eq:trueTFB}) describe the presence of a sharp step between left reservoir and intermediate plateau -- representing the shockwave front -- while the second rows describe the rarefaction wave (RW) previously depicted in figure \ref{fig:damBreakProblems}. In the perturbative regime the left and right boundaries of the RW wave become almost equal so that the RW itself becomes a nearly vertical step like those previously seen in figure \ref{fig:TFplot}. 

In the case of a dry-channel dam break, equations (\ref{eq:trueTFA}) and (\ref{eq:trueTFB}) are modified by removing rows three, while the left boundary of rows two and the right boundary of rows four are simultaneously replaced by $-2c_{1}t$.

\section{Integral-Airy function solutions and the linearized Korteweg-de Vries equation}
\label{sec:appendixkdv}

In Section \ref{sec:asymptotic}, the perturbative dam break solution at long times is shown to be given by the integral of an Airy function. These are known to be solutions of the linearized Korteweg-de Vries (KdV) equation which is first order in time and third order in space \cite{Washimi_1966,Gurevich_1973,Coutant_2014,Berry_2019,Kamchatnov_2021}. Indeed, the integral-Airy solution (\ref{eq:airyInt}) could have been alternatively derived by canonically quantizing the third order approximate dispersion relation (\ref{eq:approxDisp}) in the same manner as equation (\ref{eq:waveEqn}). This results in two uncoupled, first order in time, linear partial differential equations
\begin{equation}
    \bigg(\partial_{t}\pm c_{0}\partial_{x}\mp\frac{c_{0}\xi_{0}^{2}}{8}\partial_{x}^{3}\bigg)\eta_{\pm}(x,t)=0 \ . \label{eq:KdV}
\end{equation}
These can be further simplified by the substitution of variables $x\rightarrow 2 X\mp c_{0}t$ which yields the simplest form of linear KdV equation
\begin{equation}
    \bigg(\partial_{t}\pm c_{0}\xi_{0}^{2}\partial_{X}^{3}\bigg)\eta_{\pm}(z,t)=0 \ . \label{eq:KdV2}
\end{equation}
Solving (\ref{eq:KdV}) via Fourier transforms and re-substituting in the definition of $X$ to regain the $x$ variable precisely gives our integral-Airy function solution (\ref{eq:airyInt}). 

\section{Optical rainbows}
\label{sec:appendixrainbow}

\begin{figure}[!t]
	\centering
	\begin{minipage}{0.5\textwidth}
		\centering
		\includegraphics[width=6cm]{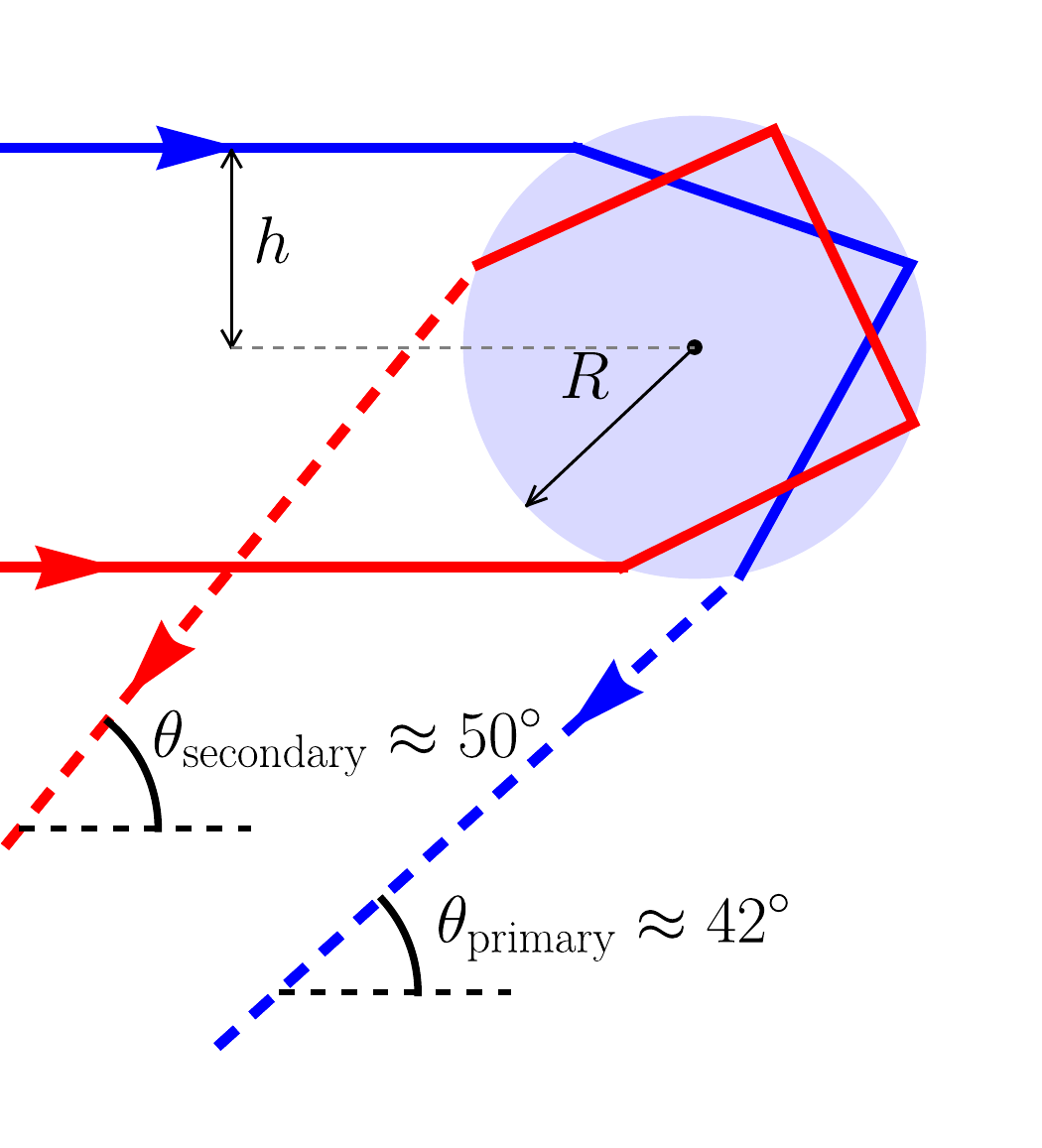}
	\end{minipage}%
	\begin{minipage}{0.5\textwidth}
		\centering
        \includegraphics[width=6cm]{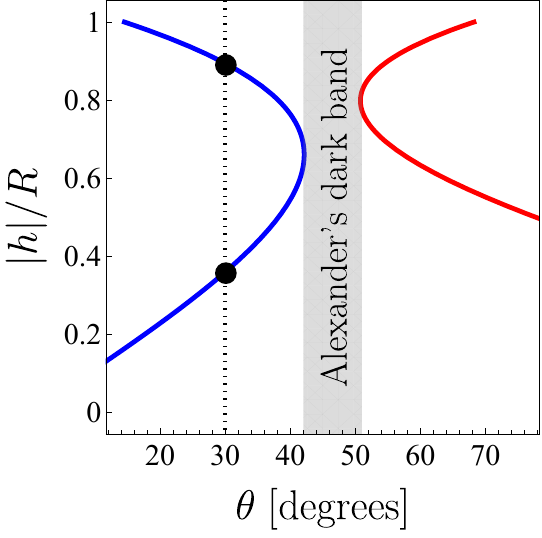}
	\end{minipage}\\[-0.2cm]	
    \caption{\textbf{Left:} Schematic diagram showing two of the Sun's rays being scattered by a water droplet (represented as a 2D circle of radius $R$) and contributing to a  double rainbow.
    The behaviour of each ray is determined by its input height $h$. The blue ray contributes to the primary rainbow and has a single internal reflection whereas the red ray undergoes two internal reflections and contributes to the secondary rainbow. The input heights we have chosen for the blue and red rays are such that they are actually the rays that output precisely at the primary and secondary rainbow angles, $\theta_{\mathrm{primary}}$ and $\theta_{\mathrm{secondary}}$, respectively, and hence correspond to the two caustics. An observer on the ground will see the brightest intensity at the two rainbow angles.
     \textbf{Right:} Plot of ray input height magnitude $\vert h\vert$ normalized by $R$ versus output ray angle $\theta$. The right and leftmost points of the blue and red curves respectively correspond to $\theta_{\mathrm{primary}}$ and $\theta_{\mathrm{secondary}}$, the region existing between them being Alexander's dark band.}\label{fig:rainbowRays}
\end{figure}

Consider the case of a single optical rainbow, i.e.\ the primary bow shown in figure \ref{fig:doubleRainbow}. This is formed when parallel light rays from the Sun enter water droplets, and are internally reflected once before leaving. To understand the phenomenon we need only consider a vertical slice through the rainbow (rather than the full bow which an observer on the ground sees due to off-axis droplets) and, indeed, we need only consider a vertical slice through a single droplet. Geometric optics shows that light scattered into each output angle $\theta$ is made up of two input rays that enter the droplet at two different heights $h$. This is illustrated in the right hand panel of figure \ref{fig:rainbowRays} by the two points at which the vertical dotted black line  crosses the blue input-output curve, indicating the two values of $h$ that contribute to the output angle $\theta=30^{\circ}$. Sliding the vertical line to the right, one sees that the multi-valued input-output relation has a stationary point where $d \theta /d h=0$. This gives rise to an angular caustic where rays coalesce and produces a focusing effect because to first order there is no change in the output angle for an infinitesimal range of rays entering at nearby heights. For the primary bow this leads to a sharp peak in the intensity at an angle of $\theta_{\mathrm{primary}} \approx 42^{\circ}$ above the horizontal. The left hand panel of figure \ref{fig:rainbowRays} shows two examples of ray trajectories in a water droplet, and in fact we chose the ones that output at the caustics for the primary and secondary bows. For an image displaying the more complete ray structure, see figure 5 in reference \cite{Link_2024}.

A double rainbow consisting of the two bows seen in figure \ref{fig:doubleRainbow} occurs because rather than exiting after a single internal reflection, the rays can instead exit after two reflections, as shown by the red ray and the corresponding red input-output  curve in figure \ref{fig:rainbowRays}. Although even higher order rainbows can exist due to additional internal reflections, they become increasingly faint and we do not consider them here. The input-output relation for the second rainbow included in figure \ref{fig:rainbowRays} lies to the right of that for the first  and is inverted. It gives rise to a ray-coalescence (caustic) at $\theta_{\mathrm{secondary}} \approx 50^{\circ}$. The key point is that the double rainbow has qualitatively the same total ray structure as that underlying the dispersive waves created in a dam break as shown in the top left panel in figure \ref{fig:doubleRainbow}, including Alexander's dark band that lies between the turning points of the two input-output relations.

\printbibliography

\end{document}